\documentclass[lettersize,journal]{IEEEtran}
\IEEEoverridecommandlockouts
\usepackage{amsmath,amsfonts}
\usepackage{amssymb}
\usepackage[caption=false,font=footnotesize,labelfont=sf,textfont=sf]{subfig}
\usepackage{stfloats}
\usepackage{url}
\usepackage{graphicx}
\usepackage{cite}
\usepackage{float}
\usepackage{subfig}
\usepackage{multirow}
\usepackage{color}
\usepackage{makecell}
\usepackage{booktabs}
\usepackage{mathtools}
\usepackage[colorlinks,citecolor=blue,linkcolor=blue]{hyperref}
\newtheorem{corollary}{Corollary}

\newtheorem{theorem}{Theorem}

\newtheorem{property}{Property}
\newtheorem{definition}{Definition}
\newtheorem{algorithm}{Algorithm}
\usepackage{algorithm, algorithmicx, algpseudocode}
\begin{document}

\title{Cutoff Rate Bounds and Constellation Shaping under Mixed Gaussian-Impulsive Noise}

\author{Tianfu Qi, \emph{Graduate Student Member, IEEE}, Jun Wang, \emph{Senior Member, IEEE}
\thanks{Tianfu Qi, Jun Wang are with the National Key Laboratory on Wireless Communications, University of Electronic Science and Technology of China, Chengdu 611731, China (e-mail: 202311220634@std.uestc.edu.cn).}
}

\maketitle

\begin{abstract}
Mixed noise, composed of white Gaussian noise (WGN) and impulsive noise (IN), appears in numerous communication scenarios and can severely degrade system performance. In this paper, we optimize the transmitted constellation under mixed noise based on a theoretical analysis of the cutoff rate (CR). First, starting from the passband model of the mixed noise, we derive its corresponding baseband representation. Then, the baseband model is employed to obtain closed-form lower and upper bounds of the CR. A piecewise linear approximation is applied to derive efficient bounds by exploiting the algebraic properties of the integral terms. These bounds are then used as criteria to optimize the transmitted constellation points in both geometric positions and probabilistic distributions. The projected gradient method is employed to solve the optimization problem, and the convergence and properties of the solutions are analyzed. Numerical results demonstrate that the proposed CR bounds are tight and exhibit the expected asymptotic behavior. Furthermore, the optimized constellation scheme achieves a significant rate improvement compared to baselines. 
\end{abstract}


\section{Introduction}
Impulsive noise (IN) appears in many practical communication scenarios, including underwater acoustic communication \cite{paper1,paper2}, atmospheric noise \cite{paper3}, NOMA systems \cite{paper4}, IoT wireless networks \cite{paper5}, smart grid communications \cite{paper6}, and wideband powerline communication \cite{paper7}. The main characteristic of IN is its large amplitude over very short durations. Combined with the unavoidable white Gaussian noise (WGN), the mixed channel noise arises. Ignoring IN and applying conventional algorithms designed for WGN scenarios can lead to severe performance degradation. To address this issue, accurate noise modeling is essential, providing a foundation for the design of effective signal processing techniques.

Recently, numerous studies have focused on developing statistical models for pure IN. For example, Middleton class A/B/C (MCA/B/C) models, proposed in \cite{paper35}, are based on the physical characteristics and spectral properties of the noise. These models are heavy-tailed distributions that can accurately capture non-Gaussian IN across various scenarios, but their probability density functions (PDFs) are quite complex. Shao et al. employed the $\alpha$-stable distribution as an approximate model for IN, which has also become one of the most widely used approaches. However, the $\alpha$-stable distribution lacks an analytic PDF except for special cases, such as the Cauchy distribution ($\alpha=1$) and Gaussian distribution ($\alpha=2$) \cite{paper8}, and it does not account for the presence of WGN. In our previous work \cite{paper9,paper10}, we proposed the GS noise model, defined as a weighted sum of a multivariate Gaussian distribution and student distribution, to characterize mixed noise with and without memory. This model admits a simple PDF, and experimental results demonstrate that it accurately represents real mixed-channel noise, enabling both theoretical analysis and practical applications.

To enhance communication system performance, receiver-side signal processing algorithms must be carefully designed, as studied extensively in the literature \cite{paper11,paper12,paper13,paper14,paper15,paper16,paper17,paper18}. Beyond receiver design, an equally powerful approach is to optimize the transmitted signal. Standard modulation schemes, such as square QAM constellations, are generally suboptimal. In WGN scenarios, it has been shown that discretizing a continuous Gaussian distribution can yield asymptotically optimal constellation points \cite{paper24,paper25}. However, practical systems with finite modulation orders remain suboptimal. Consequently, constellation optimization in more complex scenarios is typically formulated as an optimization problem and solved using iterative algorithms. Commonly used criteria include maximizing mutual information (MI), cutoff rate (CR) and ergodic capacity, as well as minimizing detection error probabilities \cite{paper26,paper27}. Many studies have investigated constellation shaping under WGN channels in various applications, including underwater wireless optical communication \cite{paper19}, ambient IoT networks \cite{paper20}, integrated sensing and communication (ISAC) systems \cite{paper21}, visible light communications \cite{paper22}, and covert communications \cite{paper23}.

However, to the best of the authors' knowledge, few existing works have addressed the optimization of constellation points under non-Gaussian mixed channel noise. The challenges arise from two main aspects. First, the statistical model and corresponding PDF for mixed noise are highly complex, a problem addressed in our previous work. More importantly, the performance metrics differ from those in WGN scenarios and are difficult to express in closed form. Minimizing Euclidean distance is generally not optimal under mixed noise. Additionally, calculating rate-based criteria typically requires high-resolution numerical integration at each iteration, imposing a prohibitive computational burden for practical optimization. Therefore, accurate analytic bounds or general distance-based criteria are essential.

Based on these motivations, the main contributions of this work can be summarized as follows,
\begin{itemize}
\item{\textbf{Theoretical perspectives:} First, to eliminate the influence of the carrier in subsequent analysis, we examine the baseband model of mixed channel noise derived from its passband representation. It is shown that the in-phase and quadrature components are uncorrelated but not independent, and their joint distribution corresponds to the bivariate form of the passband mixed noise model. Furthermore, according to the baseband model, we derive lower and upper bounds of the CR. The problem is reduced to evaluating three types of integrals. However, the second and third types do not admit closed-form solutions. To address this problem, we further simplify the integrals by analyzing the convexity and applying a piecewise linear approximation. Specifically, piecewise linear functions are used to bound certain integrand terms based on an analysis of their algebraic properties, including zero behaviors. In this way, all derived bounds are analytic and converge to the exact value as the signal-to-noise ratio (SNR) increases.}
\item{\textbf{Practical aspects:} We use the CR lower bound as the objective function to optimize the transmitted constellation points. The overall optimization problem is divided into probabilistic shaping and geometric shaping. We prove that the loss function is $L$-smooth and derive the explicit Lipschitz constant. Based on these results, a projected gradient descent (PGD) method is employed to iteratively solve the two subproblems. Additionally, the convergence and properties of the solutions are analyzed. Numerical results demonstrate that the proposed bounds are tight and exhibit the expected asymptotic behavior. Moreover, the optimized constellation scheme achieves the highest CR among baseline designs, particularly in low- and medium-SNR regimes.}
\end{itemize}

The remainder of the paper is organized as follows. Section \ref{section_baseband_noise_model} briefly introduces the passband model of mixed noise and derives the corresponding baseband representation. Based on the baseband model, lower and upper bounds of the CR for mixed noise are analyzed in Section \ref{section_CR_bounds}. Using the lower bound, constellation shaping and convergence analysis are presented in Section \ref{section_constellation_shaping}. Simulation results are provided in Section \ref{section_simulations}, and finally, we conclude this paper in Section \ref{section_conclusions}.

\textbf{Notations:} In this paper, unbolded uppercase and lowercase letters, e.g., $X$ and $x$, denote a random variable (RV) and its realization, respectively. Bold lowercase and uppercase letters, e.g., $\mathbf{x}$ and $\mathbf{X}$, represent vectors and random vectors. $\nabla_{x}^p f(\cdot)$ denote the $p$-th order derivatives of $f(\cdot)$ with respect to $x$ where $p$ is a non-negative integer. $\Gamma(\cdot)$ denotes the Gamma function and $\text{lcm}(a,b)$ represents the least common multiple. $\mathbb{E}_{X}[\cdot]$ denotes the expectation with respect to $X$.

\section{Baseband noise model}\label{section_baseband_noise_model}
\subsection{Passband mixed noise model}
We assume that the memoryless mixed noise consists of background WGN and non-Gaussian IN, i.e.,
\begin{align}\label{bursty_mixed_noise_model}
N=N_G+N_I,
\end{align}
where $N$, $N_G$, and $N_I$ denote the RVs of the mixed noise, WGN and IN, respectively. Moreover, $N_G$ and $N_I$ are typically assumed to be mutually independent, which is reasonable since they originate from different sources. The IN component can be accurately modeled by the $\alpha$-stable distribution \cite{paper8}. Furthermore, we assume that the noise has zero mean. Under this assumption, the amplitude of $N_I$ follows a symmetric $\alpha$-stable (S$\alpha$S) distribution. However, $N$ does not have a closed-form PDF, causing significant challenges for practical applications. In our prior work \cite{paper9}, an approximate PDF for $N$ (referred to `GS model') was proposed as follows,
\begin{align}\label{GS_model}
f_{N}(n)=\rho k_1\exp\left(-\frac{n^2}{4\gamma_g^2}\right) +\frac{(1-\rho)k_2}{(1+n^2/(2\alpha\gamma_s^2))^\frac{\alpha+1}{2}},
\end{align}
where $k_1$ and $k_2$ are normalization factors, given by $k_1=(2\sqrt{\pi}\gamma_g)^{-1}$, $k_2=\Gamma((\alpha+1)/2)/(\Gamma(\alpha/2)\sqrt{2\alpha\pi\gamma_s^2})$. We now describe the parameters in \eqref{GS_model}, including $\alpha$, $\gamma_g$, $\gamma_s$ and $\rho$. Similar to the S$\alpha$S distribution, $\alpha$ is the characteristic parameter representing the heaviness of the PDF tail. For practical communication applications, we consider $\alpha \in (0,2)$, which corresponds to the $\alpha$-stable distribution. $\gamma_g \in (0,+\infty)$ and $\gamma_s \in (0,+\infty)$ are the scale parameters of WGN and IN, respectively. $\rho \in [0,1]$ is a weighting parameter that adjusts the mainlobe and tail of the PDF, representing the proportion of WGN and memoryless IN. It should be noted that the complete GS model in \cite{paper9} is a multivariate distribution, primarily used to model mixed noise with or without memory. Hence, \eqref{GS_model} represents a special case of the GS noise model, reducing to the one-dimensional scenario. In this paper, we focus on memoryless mixed noise, and when referring to the `GS model', we mean the PDF given in \eqref{GS_model}.

In \cite{paper9}, we demonstrated that the GS model offers several advantages, which can be briefly summarized as follows,
\begin{itemize}
\item{The GS noise model can accurately capture the statistical properties of real memoryless mixed noise. Moreover, it provides a PDF in closed form, which is highly useful for algorithm design and theoretical analysis.}
\item{The GS model is general as it can degenerate to simpler noise models such as WGN, pure IN and Cauchy noise. In addition, an efficient parameter estimation algorithm is available in \cite{paper9}.}
\end{itemize}

\subsection{Baseband model}
In this paper, the constellation points to be optimized are two-dimensional. Therefore, the baseband noise model can be used to simplify following analyses by eliminating the influence of the carrier. Without loss of generality, we consider the passband channel noise undergoing frequency shifting and low-pass filtering at the receiver. For WGN scenarios, it is straightforward to show that after these operations, the resulting in-phase and quadrature noise components remain mutually independent and Gaussian distributed. In general, however, this property does not hold for non-Gaussian noise. For instance, under the MCA model, the in-phase and quadrature components are uncorrelated but not independent \cite{paper34}. Consequently, it is necessary to derive the explicit distribution of the baseband noise based on the GS noise model.

Let the baseband in-phase and quadrature noise at the $k$-th time instant be denoted by $N^I(k)$ and $N^Q(k)$, respectively. Then, we have
\begin{align}
N^I(k)=\sum_{m=1}^{M}h(m)N(m-k)\cos\Big(2\pi \frac{f_c}{f_s}(m-k)\Big),
\end{align}
where $f_c$ and $f_s$ separately denote the carrier frequency and sampling rate. $h(\cdot)$ represents the coefficients of the low-pass filter with the order $M_h$. The quadrature component is denoted similarly. Together with \eqref{bursty_mixed_noise_model},
\begin{align}\label{processing_procedure}
N^I(k)=&\underbrace{\sum_{m=1}^{M_h}h(m)N_G(m-k)\cos\Big(2\pi \frac{f_c}{f_s}(m-k)\Big)}_{N_G^I(k)}\nonumber\\
&+\underbrace{\sum_{m=1}^{M_h}h(m)N_I(m-k)\cos\Big(2\pi \frac{f_c}{f_s}(m-k)\Big)}_{N_I^I(k)}.
\end{align}

Based on the linearity property of stable-distributed random variables, it is straightforward that $N_G^I(k)$ and $N_I^I(k)$ still follow Gaussian and S$\alpha$S distributions, respectively. Next, the transformed parameters need to be derived. For $N_G^I(k)$, its characteristic function (CF) is given by
\begin{align}
\phi_{N_G^I(k)}(t)=\exp\bigg(-\gamma_g^2t^2\sum_{m=1}^{M_h}h(m)^2\cos\Big(2\pi\frac{f_c}{f_s}(m-k)\Big)^2\bigg).
\end{align}

Note that the coefficient sequence $\{h(\cdot)^2\}$ still forms a low-pass filter with a narrower passband. Therefore, the CF of $N_G^I(k)$ can be further simplified as $\phi_{N_G^I(k)}(t) = \exp(-\gamma_g^2 t^2 B / 4)$, where $B$ denotes the passband of the low-pass filter defined by $\{h(\cdot)\}$. For $N_I^I(k)$, the CF cannot be simplified in the same way as $N_G^I(k)$, since the $\alpha$-th order moment cannot be decomposed into multiple components. Let the characteristic and scale parameters of $N_I$ in \eqref{bursty_mixed_noise_model} be $\alpha$ and $\gamma$, respectively. Note that $\gamma$ does not necessarily equal $\gamma_s$ in \eqref{GS_model}. Then, the CF can still be expressed as follows,
\begin{align}
&\phi_{N_I^I(k)}(t)\nonumber\\
=&\exp\bigg(-\gamma|t|^{\alpha}\underbrace{\sum_{m=1}^{M_h}\Big|h(m)\cos\Big(2\pi\frac{f_c}{f_s}(m-k)\Big)\Big|^{\alpha}}_{\triangleq p(k,M_h)}\bigg).\nonumber\\
\end{align}

It can be observed that $p(k,M_h)$ is equivalent to the convolution of a periodic signal with the transformed low-pass filter coefficient sequence, namely,
\begin{align}
p(k,M_h)=&(p_1\circledast p_2)(k,M_h)\nonumber\\
\triangleq&\sum_{m=1}^{M_h}p_1(m)p_2(m-k),
\end{align}
where $p_1(m)\triangleq |h(m)|^{\alpha}$ and $p_2(m)\triangleq|\cos(2\pi f_cm/f_s)|^{\alpha}$. It can be shown that $p_1(m)$ corresponds to a low-pass filter in the frequency domain. Consequently, $p(k,M_h)$ reduces to a constant related to the power of the carrier signal and the sequence $\{h(\cdot)\}$. We omit the time index of $p(k,M_h)$ in the following due to its time-invariant property. Based on Parseval's theorem, it follows that
\begin{align}\label{power_of_processed_IN}
p(M_h)=\frac{1}{\text{lcm}(f_s,f_c)}\sum_{m=1}^{M_h}|h(m)|^{\alpha}\sum_{m=1}^{\text{lcm}(f_s,f_c)}\Big|\cos\Big(2\pi\frac{f_c}{f_s}m\Big)\Big|^{\alpha}.
\end{align}

The above analysis focuses on the marginal distributions of $N^I(k)$ and $N^Q(k)$ and their corresponding parameter transformations. Next, we consider the joint distribution of $N^I(k)$ and $N^Q(k)$. We first express them in vector form, i.e.,
\begin{align}\label{GS_form}
\mathbf{N}(k)\triangleq&[N^I(k),N^Q(k)]\nonumber\\
=&[N^I_G(k),N^Q_G(k)]+[N^I_I(k),N^Q_I(k)].
\end{align}

Clearly, $[N^I_G(k), N^Q_G(k)]$ forms a bivariate Gaussian distribution with an identity correlation matrix. For any S$\alpha$S-distributed random variable $X$, it can be represented as the product of a Gaussian RV and the square root of a positive, right-skewed $\alpha$-stable RV, as follows,
\begin{align}\label{aSG_decomposition}
N_I(k)=\sqrt{X(k)}G(k),
\end{align}
where $X(k)\sim \mathcal{S}(\alpha/2,1,(\cos(4\pi/\alpha))^{2/\alpha},0)$ and $G(k)$ follows the Gaussian distribution where $\mathcal{S}(\cdot)$ denotes the $\alpha$-stable distribution. Combining \eqref{processing_procedure} and \eqref{aSG_decomposition}, the joint distribution of $[N^I_I(k),N^Q_I(k)]$ is equivalent to the analysis for $[N^I_G(k),N^Q_G(k)]$. For example, we have
\begin{align}\label{IQ_IN}
[N^I_I(k),N^Q_I(k)]=\sqrt{X(k)}[G^I(k),G^Q(k)],
\end{align}
where the covariance matrix of $[G^I(k), G^Q(k)]$ can be analyzed via its CF, and it remains a scaled identity matrix. Finally, using \eqref{GS_form}, $[N^I(k), N^Q(k)]$ follows the standard form of bursty mixed noise described in \cite{paper9}. This property indicates that the joint distribution of $[N^I(k), N^Q(k)]$ can be accurately modeled by the GS noise model, expressed as follows,
\begin{align}\label{baseband_PDF}
f_{\mathbf{N}}(\mathbf{n})=\rho k_3\exp\left(-\frac{\Vert\mathbf{n}\Vert^2}{4\gamma_g^2}\right) +\frac{(1-\rho)k_4}{(1+\Vert\mathbf{n}\Vert^2/(2\alpha\gamma_s^2))^\frac{\alpha+1}{2}},
\end{align}
where $k_3=(2\sqrt{\pi}\gamma_g)^{-2}$ and $k_4=\Gamma((\alpha+2)/2)/(\sqrt{2}\Gamma(\alpha/2)\alpha\pi\gamma_s)$. We omit the time index in \eqref{baseband_PDF} since the noise parameters are time-invariant. $f_{\mathbf{N}}(\mathbf{n})$ represents the baseband model for memoryless mixed noise, which will be used throughout the remainder of the paper. It should be noted that although $f_{\mathbf{N}}(\mathbf{n})$ follows the complete GS noise model, there is no memory between the in-phase and quadrature components as the correlation matrix is diagonal. Furthermore, with a slight abuse of notation, the parameters in \eqref{baseband_PDF} are not identical to those in \eqref{GS_model}.



\section{Bounds of CR}\label{section_CR_bounds}
In this section, we derive the CR under memoryless mixed channel noise. However, the exact expressions are generally too complex to be obtained in closed form. Therefore, we focus on deriving analytic lower and upper bounds.

Before proceeding, the baseband received signal model is given as follows,
\begin{align}\label{received_signal_model}
\mathbf{y}(k)=\mathbf{s}(k)+\mathbf{n}(k),k=1,2,\cdots,
\end{align}
where $\mathbf{y}(k)$, $\mathbf{s}(k)$ and $\mathbf{n}(k)$ separately denote the received signal vector, transmitted constellation vector and baseband mixed noise vector at the $k$-th time instant. The PDF of $\mathbf{n}(k)$ is given in \eqref{baseband_PDF}, and $\mathbf{s}(k) \in \Omega_s$ where $\Omega_s$ denotes the set of feasible constellation points.

As a measure of the maximum achievable rate for reliable transmission, the CR serves as a practical performance metric for communication systems. Given transmitted constellation points $\mathbf{s}_k \in \Omega_s$ and corresponding probabilities $p_k$, $k=1,\dots,|\Omega_s|$, the CR is expressed as \cite{paper33}
\begin{align}\label{CR_original}
R(\Omega_s,\Theta)\triangleq&-\log\bigg(\int_{\mathcal{R}^2}\bigg(\sum_{k=1}^{|\Omega_s|}p_k\sqrt{f_{\mathbf{N}}(\mathbf{y}-\mathbf{s}_k)}\bigg)^2\bigg)d\mathbf{y}\nonumber\\
=&-\log\bigg(\sum_{k=1}^{|\Omega_s|}p_k^2+\underbrace{\sum_{k=1}^{|\Omega_s|-1}\sum_{l=k+1}^{|\Omega_s|}p_kp_lZ(k,l)}_{\triangleq\bar{Z}}\bigg),
\end{align}
where
\begin{align}\label{Z_definition}
Z(k,l)\triangleq\int_{\mathcal{R}^2}\sqrt{f_{\mathbf{N}}(\mathbf{y}-\mathbf{s}_k)f_{\mathbf{N}}(\mathbf{y}-\mathbf{s}_l)}d\mathbf{y}.
\end{align}

In \eqref{CR_original}, $\Theta$ denotes the set of channel noise parameters. Note that the subscript $k$ refers to the $k$-th element in the set $\Omega_s$ instead of the time index. $Z(k,l)$, also known as the Bhattacharyya parameter, is an upper bound of detection error probability assuming only $\mathbf{s}_k$ and $\mathbf{s}_l$ are transmitted with uniform probability. The remaining task is to derive an analytic expression for $Z(k,l)$. Unfortunately, this is extremely challenging because the PDF is a weighted sum of components. In this case, the Gaussian mainlobe and the heavy-tail part are coupled and no closed-form expression exists even with the use of special functions. Therefore, deriving lower and upper bounds for $Z(k,l)$ is essential to facilitate further applications.

\subsection{Lower Bounds for CR}\label{lower_bound_for_CR}
Deriving a lower bound for the CR is equivalent to establishing an upper bound for $Z(k,l)$. For clarity, we first introduce the following notations,
\begin{align}
\Delta\mathbf{s}(k,l)\triangleq\mathbf{s}_k-\mathbf{s}_l,
\end{align}
\begin{align}
S_1(k,l)\triangleq \rho k_3\exp\left(-\frac{\Vert\mathbf{n}\Vert^2}{8\gamma_g^2}\right)\exp\left(-\frac{\Vert\mathbf{n}-\Delta\mathbf{s}(k,l)\Vert^2}{8\gamma_g^2}\right),
\end{align}
\begin{align}
S_2(k,l)\triangleq&\sqrt{\rho(1-\rho)k_3k_4}\exp\Big(-\frac{\Vert\mathbf{n}\Vert^2}{8\gamma_g^2}\Big)\nonumber\\
&\times\Big(\frac{1+\Vert\mathbf{n}-\Delta\mathbf{s}(k,l)\Vert^2}{2\alpha\gamma_s^2}\Big)^{-\frac{\alpha+2}{4}},
\end{align}
\begin{align}
S_3(k,l)\triangleq& (1-\rho)k_4\Big(\Big(\frac{1+\Vert\mathbf{n}\Vert^2}{2\alpha\gamma_s^2}\Big)\nonumber\\
&\times\Big(\frac{1+\Vert\mathbf{n}-\Delta\mathbf{s}(k,l)\Vert^2}{2\alpha\gamma_s^2}\Big)\Big)^{-\frac{\alpha+2}{4}}.
\end{align}

For convenience, we omit indicating the integral variable in $S_j(k,l)$ with $j=1,2,3$, which are functions of the noise vector $\mathbf{n}$ by default. Then, $Z(k,l)$ can be rewritten as follows,
\begin{align}\label{Z_S_original}
Z(k,l)=\int_{\mathcal{R}^2}\sqrt{S_1(k,l)^2+2S_2(k,l)^2+S_3(k,l)^3}d\mathbf{n}.
\end{align}

Due to the square root operation, it is impossible to derive a closed-form expression for $Z(k,l)$. To address this, the integrand must first be decomposed into separate terms. By applying an elementary inequality, we obtain
\begin{align}\label{Z_upper_bound}
Z(k,l)\leq\int_{\mathcal{R}^2}S_1(k,l)+\sqrt{2}S_2(k,l)+S_3(k,l)d\mathbf{n}.
\end{align}

In this way, bounds can be separately designed for the three terms, as outlined in the following steps.

\textbf{Step 1}: The simplest case is the integral of $S_1(k,l)$ over $\mathcal{R}^2$, and an exact expression can be derived using a coordinate transformation. For instance,
\begin{align}
\int_{\mathcal{R}^2}S_1(k,l)d\mathbf{n}=4\pi\gamma_g^2\rho k_3\exp\Big(-\frac{\Vert\Delta\mathbf{s}(k,l)\Vert^2}{16\gamma_g^2}\Big),
\end{align}
where the derivation relies on argument shifting and the Gaussian kernel integral identity.

\textbf{Step 2}: Unlike $S_1(k,l)$, the derivation of $S_2(k,l)$ and $S_3(k,l)$ is more complex due to the constellation point difference $\Delta\mathbf{s}(k,l)$, which makes the polar coordinate transformation inapplicable. A natural approach is to apply the triangle inequality to the term $\Vert\mathbf{n}-\Delta\mathbf{s}(k,l)\Vert^2$, which appears to yield both lower and upper bounds, i.e.,
\begin{align}\label{trivial_upper_bound_S_12}
\int_{\mathcal{R}^2}S_2(k,l)d\mathbf{n}\leq&\int_{\mathcal{R}^2}\Big(\frac{1+\Vert\mathbf{n}-\Delta\mathbf{s}(k,l)\Vert^2}{1+|\Vert\mathbf{n}\Vert^2-\Vert\Delta\mathbf{s}(k,l)\Vert^2|}\Big)^{\frac{\alpha+2}{4}}\nonumber\\
&\times S_2(k,l)d\mathbf{n}\nonumber\\
\leq&2\pi\sqrt{\rho(1-\rho)k_3k_4}\int_{0}^{+\infty}r\exp\Big(-\frac{r^2}{8\gamma_g^2}\Big)\nonumber\\ &\times\Big(\frac{|r^2-\Vert\Delta\mathbf{s}(k,l)\Vert^2|}{2\alpha\gamma_s^2}\Big)^{-\frac{\alpha+2}{4}}dr.
\end{align}

The above is an improper integral but its value is finite and can be expressed in closed form. Using similar steps, the lower bound can also be derived. Unfortunately, both bounds deviate significantly from the exact value. The first issue arises from the triangle inequality since $\Vert\mathbf{n}-\Delta\mathbf{s}(k,l)\Vert$ equals zero only when $\mathbf{n}=\Delta\mathbf{s}(k,l)$. After applying the inequality, the condition relaxes to $\Vert\mathbf{n}\Vert=\Vert\Delta\mathbf{s}(k,l)\Vert$, which is much weaker. Then, the constant ignorance makes the integral trackable but improper. As a result, these bounds may deviate significantly from the accurate value by nearly two orders of magnitude. This motivates the need for more refined techniques to construct a tighter upper bound.

To this end, we apply the Given's transform to $\mathbf{n}$ such that $\Delta\mathbf{s}(k,l)$ is paralleled to the axis corresponding to $n_1$. In this way, there is no shift in the $n_2$ component and therefore, we first compute the inner integral with respect to $n_2$, i.e.,
\begin{align}\label{S_2_med_1}
&\int_{\mathcal{R}^2}S_2(k,l)d\mathbf{n}\nonumber\\
=&\delta_1\int_{-\infty}^{+\infty}\exp\Big(-\frac{n_1^2}{8\gamma_g^2}\Big)\int_{0}^{+\infty}\exp\Big(-\frac{n_2^2}{8\gamma_g^2}\Big)\nonumber\\
&\times(2\alpha\gamma_s^2+(n_1-\Vert\Delta\mathbf{s}(k,l)\Vert)^2+n_2^2)^{-\frac{\alpha+2}{4}}dn_2dn_1\nonumber\\
\overset{(a)}{=}&\frac{\delta_1\Gamma(1/2)}{2}\int_{-\infty}^{+\infty}(2\alpha\gamma_s^2+n_1^2)^{-\frac{\alpha}{4}}\Psi\Big(\frac{1}{2},1-\frac{\alpha}{4};\frac{2\alpha\gamma_s^2+n_1^2}{8\gamma_g^2}\Big)\nonumber\\
&\times\exp\Big(-\frac{(n_1+\Vert\Delta\mathbf{s}(k,l)\Vert)^2}{8\gamma_g^2}\Big)dn_1,
\end{align}
where $\delta_1=2\sqrt{\rho(1-\rho)k_3k_4}(2\alpha\gamma_s^2)^{\frac{\alpha+2}{4}}$ and $\Psi(\cdot)$ denotes the Tricomi hypergeometric function. The $(a)$ is due to the following identity \cite{book1},
\begin{align}
\int_{0}^{+\infty}\frac{x^{a-1}}{(x+z)^b}\exp(-px)dx=\Gamma(a)z^{a-b}\Psi(a,a-b+1;pz).
\end{align}

The expression in \eqref{S_2_med_1} cannot be further simplified using standard transformations or known integral identities. To address this challenge, we first define
\begin{align}
J_1(x)\triangleq&(2\alpha\gamma_s^2+x^2)^{-\frac{\alpha}{4}},\\
J_2(x)\triangleq&\Psi\Big(\frac{1}{2},1-\frac{\alpha}{4};\frac{2\alpha\gamma_s^2+x^2}{8\gamma_g^2}\Big).
\end{align}

We observe that the argument of $J_2(x)$ is essentially a scaled version of $2\alpha\gamma_s^2+x^2$. Thus, $J_1(x)$ and $J_2(x)$ are expected to share similar structural properties such as monotonicity. In addition, the slope of $J_1(x)$ does not vary abruptly with respect to $x$. Its convexity can also be directly verified. For instance, we have
\begin{align}\label{convexity_solution}
\nabla^2_xJ_1(x)=0\Rightarrow x=\pm\sqrt{\frac{4\alpha\gamma_s^2}{\alpha+2}}.
\end{align}

Based on these advantages, it is efficient to use a linear function as the upper and lower bounds for $J_1(x)$. In other words, the piecewise linear approximation (PLA) can be applied to both $J_1(x)$ and $J_2(x)$. Substituting these approximations into \eqref{S_2_med_1} significantly reduces the complexity of obtaining an analytical expression. To implement this approach, the convexity of $J_2(x)$ also needs to be considered. However, an analytical solution for $\nabla^2_x J_2(x)=0$ generally does not exist. In fact, rather than bounding $J_1(x)$ and $J_2(x)$ separately, we opt to use linear functions to directly bound $J_1(x)J_2(x)$ because the solution to $\nabla^2_x (J_1(x)J_2(x))=0$ closely approximates that of $\nabla^2_x J_1(x)=0$, as demonstrated in the following property.
\begin{property}\label{property_1}
\textit{The zeros of $\nabla_{x}^2 J_1(x)$ and $\nabla_{x}^2 \big(J_1(x)J_2(x)\big)$ are symmetric about the origin. Furthermore, the approximate solutions for the zeros $x^*$ of $\nabla_{x}^2 \big(J_1(x)J_2(x)\big)$ are given by
\begin{align}\label{S_2_real_convexity_zeros}
x^*=x_0-\frac{2\nabla_{x}J_1(x_0)\nabla_{x}J_2(z_0)-J_1(x_0)\nabla_{x}^2J_2(x_0)}{\nabla_{x}^3J_1(x_0)J_2(z_0)},
\end{align}
where $z_0\triangleq(2\alpha\gamma_s^2+x_0^2)/(8\gamma_g^2)$ and $x_0$ is the zeros of $\nabla_{x}^2J_1(x)$, which are shown in \eqref{convexity_solution}.}
\end{property}
\begin{IEEEproof}
The symmetry can be straightforwardly verified as they are both even functions. The zeros of $\nabla_{x}^2 \big(J_1(x)J_2(x)\big)$ can then be approximately obtained using a Taylor expansion. Specifically, a second-order expansion of $J_1(x)J_2(x)$ around $x_0$ is performed, leading directly to \eqref{S_2_real_convexity_zeros}. It is important to note that this approximation is more accurate when $x^*$ is close to $x_0$, which holds in the present case. In fact, we have $\nabla_{x}^2(J_1(x)J_2(x))=\nabla_{x}^2J_1(x)J_2(x)+2\nabla_{x}J_1(x)\nabla_{x}J_2(x)+J_1(x)\nabla_{x}^2J_2(x)$. The slope of the mainlobe of $J_1(x)$ varies relatively fast and at the convexity inflection point, $\nabla_{x}^2 J_1(x)$ changes rapidly. Moreover, based on properties of the Tricomi hypergeometric function, $J_2(x)$ decreases monotonically for $x \in [0, +\infty]$ with $J_2(+\infty)=0$ and $J_2(0)<+\infty$. Therefore, the shift term $2 \nabla_{x} J_1(x) \nabla_{x} J_2(x) + J_1(x) \nabla_{x}^2 J_2(x)$ is relatively small and does not introduce a significant bias to $x^*$ compared to $x_0$.
\end{IEEEproof}

Property \ref{property_1} provides an analytical expression for the zeros of $\nabla_{x}^2 \big(J_1(x)J_2(x)\big)$, which can be used to define horizontal axis divisions in PLA. In practice, it is often sufficient to directly apply \eqref{convexity_solution}, as it is efficient in most scenarios and eliminates the need for numerical methods.

Next, we introduce the following definition to facilitate the analysis.
\begin{definition}\label{definition_1}
\textit{Let $g:\mathcal{R}\rightarrow\mathcal{R}$ be a bounded function. Then, define that
\begin{align}
\mathcal{T}_U([a,b],g,x)\triangleq\{f(x):&f(x)=px+q,\forall x\in[a,b],\nonumber\\
&f(x)\geq g(x),\forall p,q\in\mathcal{R}\},
\end{align}
\begin{align}
\mathcal{T}_L([a,b],g,x)\triangleq\{f(x):&f(x)=px+q,\forall x\in[a,b],\nonumber\\
&f(x)\leq g(x),\forall p,q\in\mathcal{R}\}.
\end{align}}
\end{definition}

Definition \ref{definition_1} introduces a family of upper- and lower-bounded linear functions for any given function $g(\cdot)$ and range $[a,b]$. For $\mathcal{T}_U([a,b],g,x)$, if $g(x)$ is convex, the primary focus can be placed on the minimal linear function in the family since it achieves the smallest approximation error. Moreover, its explicit expression can be specified as follows,
\begin{align}
\inf_{f}\mathcal{T}_U([a,b],g,x)=\frac{g(b)-g(a)}{b-a}x+\frac{bg(a)-ag(b)}{b-a}.
\end{align}

When $g(x)$ is concave, a maximum element generally does not exist unless additional criteria are imposed. Here, the maximum element $f^*(x)$ is defined such that $f^*(x) \geq f(x)$ for all $f(x) \in \mathcal{T}_L([a,b],g,x)$ and $x \in [a,b]$. For example, if we introduce the mean square error criterion, namely,
\begin{align}
\min_{p,q}\frac{1}{b-a}\int_{a}^{b}(\mathcal{T}_U([a,b],g,x)-g(x))^2dx\nonumber,
\end{align}
the maximum element under this criterion is unique. However, solving for $p$ and $q$ may be challenging. Therefore, one can simply select an element from $\mathcal{T}_U([a,b],g,x)$ to simplify the derivation. A similar analysis applies to $\mathcal{T}_L([a,b],g,x)$.

We are now ready to derive a closed-form expression for \eqref{S_2_med_1}. Leveraging Definition \ref{definition_1} and the preceding analysis,
\begin{align}\label{S_2_med_2}
\int_{\mathcal{R}^2}&S_2(k,l)d\mathbf{n}\lessapprox\frac{\delta_1\Gamma(1/2)}{2}\sum_{h=1}^{K}\int_{d_{h-1}}^{d_h}\mathcal{T}_U([d_{h-1},d_h],\nonumber\\ &J_1(n_1)J_2(n_1),n_1)\exp\Big(-\frac{(n_1+\Vert\Delta\mathbf{s}(k,l)\Vert)^2}{8\gamma_g^2}\Big)dn_1,
\end{align}
where the inequality and approximation arise from the definition of $\mathcal{T}_U([a,b],g,x)$ and the truncation of the integral range. Here, the integral is evaluated over $[d_0, d_K]$ where $K$ denotes the total number of division points. In fact, this truncation produces a lower bound for the integral. Nevertheless, this loss is negligible because the range can be chosen sufficiently large to achieve the desired accuracy. Furthermore, we select the element from $\mathcal{T}_U([a,b],g,x)$ that is tangent to $g(x)$ at $(a+b)/2$ when $g$ is concave. For example,
\begin{align}
\mathcal{T}_U([a,b],g,x)=\nabla_xg\Big(\frac{a+b}{2}\Big)&x+g\Big(\frac{a+b}{2}\Big)\nonumber\\
&-\frac{a+b}{2}\nabla_xg\Big(\frac{a+b}{2}\Big).
\end{align}

This choice is sufficiently simple and its efficiency is also verified by numerical results. Here, we further introduce a shift in the truncation range. Let $r_0 = d_0 + \Vert \mathbf{s}_j \Vert$ and $r_K = d_K + \Vert \mathbf{s}_j \Vert$, then \eqref{S_2_med_2} becomes
\begin{align}
\int_{\mathcal{R}^2}&S_2(k,l)d\mathbf{n}\lessapprox\frac{\delta_1\Gamma(1/2)}{2}\sum_{h=1}^{K}\int_{r_{h-1}}^{r_h}\mathcal{T}_U([d_{h-1},d_h],\nonumber\\ &J_1(n_1)J_2(n_1),n_1-\Vert\Delta\mathbf{s}(k,l)\Vert)\exp\Big(-\frac{n_1^2}{8\gamma_g^2}\Big)dn_1.
\end{align}

Note that the purpose of this shift is solely to remove $\Vert \Delta \mathbf{s}(k,l) \Vert$ from the exponential function. Therefore, the approximating linear function should still be selected over the original range $[d_{h-1}, d_h]$. As a result, the upper bound expression becomes linear with respect to $\Vert \Delta \mathbf{s}(k,l) \Vert$, which greatly facilitates practical applications. Next, deriving the upper bound of \eqref{S_2_med_2} is equivalent to evaluating the integral
\begin{align}
&\int_{a}^{b}n_1^c\exp\Big(-\frac{n_1^2}{8\gamma_g^2}\Big)dn_1,c=0,1,\nonumber
\end{align}
which is trackable.
\begin{figure*}[t]
\begin{align}\label{S_2_upper_bound}
\int_{\mathcal{R}^2}S_2(k,l)d\mathbf{n}\leq&\frac{\delta_1\Gamma(1/2)}{2}\sum_{h=1}^{K}4p_{r_{h-1}}^{r_h}(J_1(n_1)J_2(n_1))\gamma_g^2\Big(\exp\Big(-\frac{r_{h-1}^2}{8\gamma_g^2}\Big)-\exp\Big(-\frac{r_{h}^2}{8\gamma_g^2}\Big)\Big)\nonumber\\
&+\sqrt{2\pi\gamma_g^2}(q_{r_{h-1}}^{r_h}(J_1(n_1)J_2(n_1))-p_{r_{h-1}}^{r_h}(J_1(n_1)J_2(n_1))\Vert\Delta\mathbf{s}(k,l)\Vert)\sqrt{2\pi\gamma_g^2}\Big(\mathcal{E}\Big(\frac{r_h}{\sqrt{8\gamma_g^2}}\Big)-\mathcal{E}\Big(\frac{r_h}{\sqrt{8\gamma_g^2}}\Big)\Big)\nonumber\\
&\triangleq U(S_2(k,l)).
\end{align}
\hrulefill
\end{figure*}
Combining the previous discussions, the final upper bound is given in \eqref{S_2_upper_bound} where $\mathcal{E}(\cdot)$ denotes the Gaussian error function. The slope and intercept of the approximating linear function, denoted by $p_a^b(g)$ and $q_a^b(g)$, are given by
\begin{equation}
p_a^b(g)\triangleq\left\{
\begin{aligned}
&\frac{g(b)-g(a)}{b-a},\nabla^2_xg\Big(\frac{a+b}{2}\Big)>0 \\
&\nabla_xg\Big(\frac{a+b}{2}\Big),\nabla^2_xg\Big(\frac{a+b}{2}\Big)<0, \\
\end{aligned}
\right.
\end{equation}
\begin{equation}
q_a^b(g)\triangleq\left\{
\begin{aligned}
&\frac{bg(a)-ag(b)}{b-a},\nabla^2_xg\Big(\frac{a+b}{2}\Big)>0 \\
&\frac{a+b}{2}-\frac{a+b}{2}\nabla_xg\Big(\frac{a+b}{2}\Big),\nabla^2_xg\Big(\frac{a+b}{2}\Big)<0. \\
\end{aligned}
\right.
\end{equation}

The division sequence along the horizontal axis is crucial for the tightness of \eqref{S_2_upper_bound}. The function $J_1(x)J_2(x)$ is symmetric and consists of two regions. In the mainlobe containing the origin, it is concave. As for the tail region, $J_1(x)J_2(x)$ is convex and the slope approaches zero as $|x| \rightarrow +\infty$. Therefore, the division resolution in the tail can be relatively low, which reduces the number of terms. In contrast, the function varies more rapidly in the mainlobe, necessitating a higher number of division points. Fortunately, $\sqrt{4\alpha\gamma_s^2/(\alpha+2)}$ is usually small and therefore, the mainlobe range is limited. In summary, the piecewise approximation does not increase the complexity of obtaining the upper bound due to the availability of closed-form expressions.

\textbf{Step 3}: Finally, we consider the integral over the product of the two tail regions. The overall approach is similar to `Step 2' and the repetitive details are omitted. With some manipulations, we obtain
\begin{align}\label{S_3_med_1}
&\int_{\mathcal{R}^2}S_3(k,l)d\mathbf{n}\nonumber\\
\leq&\delta_2\int_{r_0}^{r_K}(2\alpha\gamma_s^2+(n_1-\Vert\Delta\mathbf{s}(k,l)\Vert)^2)^{-\frac{\alpha}{4}}(2\alpha\gamma_s^2+n_1^2)^{-\frac{\alpha+2}{4}}\nonumber\\
&\times \prescript{}{2}F_1\Big(\frac{\alpha+2}{4},\frac{1}{2};\frac{\alpha+2}{2};\frac{2n_1\Vert\Delta\mathbf{s}(k,l)\Vert-\Vert\Delta\mathbf{s}(k,l)\Vert^2}{2\alpha\gamma_s^2+n_1^2}\Big)dn_1,
\end{align}
where $\delta_2=(1-\rho)k_4(2\alpha\gamma_s^2)^{\frac{\alpha+2}{4}}B(1/2,(\alpha+1)/2)$. $B(\cdot,\cdot)$ denotes the Beta function and $\prescript{}{2}F_1(\cdot,\cdot;\cdot;\cdot)$ is the Gaussian hypergeometric function. The derivation of \eqref{S_3_med_1} relies on that \cite{book1}
\begin{align}\label{identity_2}
&\int_{0}^{+\infty}x^{a-1}(1-x)^b(1-zx)^cdx\nonumber\\
=&B(a,-a-b-c)_{2}F_1(-c,a;-b-c;1-z).
\end{align}

Denote that
\begin{align}
J_3(x)\triangleq \prescript{}{2}F_1\Big(\frac{\alpha+2}{4},\frac{1}{2};\frac{\alpha+2}{2};\frac{2x\Vert\Delta\mathbf{s}(k,l)\Vert-\Vert\Delta\mathbf{s}(k,l)\Vert^2}{2\alpha\gamma_s^2+x^2}\Big),
\end{align}
and then, the explicit upper bound can be obtained in \eqref{S_3_upper_bound}
\begin{figure*}[b]
\hrulefill
\begin{align}\label{S_3_upper_bound}
&\int_{\mathcal{R}^2}S_3(k,l)d\mathbf{n}\leq\delta_2\sum_{h=1}^{K}\int_{r_{h-1}}^{r_h}(2\alpha\gamma_s^2+(n_1-\Vert\Delta\mathbf{s}(k,l)\Vert)^2)^{-\frac{\alpha}{4}}\mathcal{T}_U([r_{h-1},r_h],J_1(n_1)^{\frac{\alpha+2}{\alpha}})\mathcal{T}_U([r_{h-1},r_h],J_3(n_1))dn_1\nonumber\\
=&\delta_2\sum_{h=1}^{K}p_{r_{h-1}}^{r_h}(J_1(n_1)^{\frac{\alpha+2}{\alpha}})p_{r_{h-1}}^{r_h}(J_3(n_1))(G_2(r_h)-G_2(r_{h-1}))+(2\Vert\Delta\mathbf{s}(k,l)\Vert p_{r_{h-1}}^{r_h}(J_1(n_1)^{\frac{\alpha+2}{\alpha}})p_{r_{h-1}}^{r_h}(J_3(n_1))\nonumber\\
&+q_{r_{h-1}}^{r_h}(J_1(n_1)^{\frac{\alpha+2}{\alpha}})p_{r_{h-1}}^{r_h}(J_3(n_1))+p_{r_{h-1}}^{r_h}(J_1(n_1)^{\frac{\alpha+2}{\alpha}})q_{r_{h-1}}^{r_h}(J_3(n_1)))(G_1(r_h)-G_1(r_{h-1}))+(p_{r_{h-1}}^{r_h}(J_1(n_1)^{\frac{\alpha+2}{\alpha}})\nonumber\\ &\times\Vert\Delta\mathbf{s}(k,l)\Vert+q_{r_{h-1}}^{r_h}(J_1(n_1)^{\frac{\alpha+2}{\alpha}})+p_{r_{h-1}}^{r_h}(J_3(n_1))\Vert\Delta\mathbf{s}(k,l)\Vert+q_{r_{h-1}}^{r_h}(J_3(n_1)))(G_0(r_h)-G_0(r_{h-1}))\triangleq U(S_3(k,l)).
\end{align}
\end{figure*}
where
\begin{align}
G_p(x)\triangleq\frac{x^{p+1}}{p+1}(2\alpha\gamma_s^2)^{-\frac{\alpha}{4}}\prescript{}{2}F_1\Big(\frac{\alpha}{4},\frac{p+1}{2};\frac{p+3}{2};-\frac{x^2}{2\alpha\gamma_s^2}\Big).
\end{align}

The remaining task is the convexity analysis of $J_3(x)$. We next provide a method to determine the approximate zeros of $\nabla_x^2 J_3(x)$. First, note that the argument of the Gaussian hypergeometric function in $J_3(x)$ can be rewritten as $(2x-1)/(2\alpha\gamma_s^2/\Vert\Delta\mathbf{s}(k,l)\Vert+x^2)$. Let $z(x) = (2x-1)/(s + x^2)$ and $l(z(x)) = \prescript{}{2}F_1(a,b;c;z(x))$. Then, finding the zeros of $f(x)$ is equivalent to solving the following equation,
\begin{align}\label{solve_equation}
\frac{\nabla_{z(x)}^2l(x)}{\nabla_{z(x)}l(x)}=-\frac{\nabla_x^2z(x)}{(\nabla_xz(x))^2}.
\end{align}

The right-hand side (RHS) of \eqref{solve_equation} is a fraction and our main approximation focuses on the left-hand side (LHS). For the case $|z(x)| < 1$, the function $\prescript{}{2}F_1(a,b;c;z(x))$ can be expanded as an infinite series and terms of higher order (second order for $\nabla_{z(x)}l(x)$ and third order for $\nabla_{z(x)}^2l(x)$) in $x$ are neglected. In this way, we obtain
\begin{align}\label{approximate_1}
\frac{\nabla_{z(x)}^2l(x)}{\nabla_{z(x)}l(x)}\approx\frac{(b+1)(a+1)}{c+1}.
\end{align}

Then, \eqref{solve_equation} reduces to solving a cubic equation, which does not introduce additional computational complexity. For the case $|z(x)| \ge 1$, the series expansion is not convergent, and analytic continuation would be required. However, this approach is cumbersome, and the LHS of \eqref{solve_equation} cannot be simplified in the same way as for $|z(x)| < 1$ since $z$ appears in the denominator. Here, we directly employ the asymptotic approximation of the Gaussian hypergeometric function for the LHS of \eqref{solve_equation} when $|z(x)| \ge 1$. For example, we have $l(z(x)) \sim k(-z)^{-a}$ and then \cite{book1},
\begin{align}\label{approximate_2}
\frac{\nabla_{z(x)}^2l(x)}{\nabla_{z(x)}l(x)}\approx-\frac{(a+1)}{z(x)}.
\end{align}

Substituting the above approximation into \eqref{solve_equation} reduces it to solving a quartic equation, which also admits closed-form solutions. Note that the obtained solutions must lie within the corresponding feasible range. For instance, if \eqref{approximate_1} is applied but a solution $x^*$ does not satisfy $|z(x^*)| < 1$, it should be discarded.


Finally, by combining the above three steps with \eqref{Z_upper_bound}, the upper bound for $Z(k,l)$ and the corresponding lower bound of the CR can be obtained.

\subsection{Upper Bounds for CR}
Similar to the lower bound, we first decouple \eqref{Z_S_original} into simpler terms. Note that the square root function is concave and therefore,
\begin{align}\label{Z_lower_bound_old}
Z(k,l)\geq\frac{1}{2}\int_{\mathcal{R}^2}S_1(k,l)+\sqrt{2}S_2(k,l)+S_3(k,l)d\mathbf{n}.
\end{align}

\eqref{Z_lower_bound_old} is derived from the standard Jensen's inequality. It can be refined to achieve tighter bounds in the local region. For instance, the value of $S_j$ depends on the ratio of WGN to IN power. If the WGN is negligible and $\rho = 0$, then clearly $Z(k,l) = S_3(k,l)$ and \eqref{Z_lower_bound_old} becomes loose. We improve the tightness from two aspects. First, note that $S_3(k,l)$ decays much more slowly with respect to SNR than $S_1(k,l)$ and $S_2(k,l)$ due to the Gaussian kernel. Therefore, it is reasonable to retain only $S_3(k,l)$ as $\text{SNR} \rightarrow +\infty$, i.e.,
\begin{align}\label{Z_lower_bound_new1}
Z(k,l)\geq\int_{\mathcal{R}^2}S_3(k,l)d\mathbf{n}.
\end{align}

This operation introduces a relatively large error in the low-SNR region. As a complement, we employ a weighted version of Jensen's inequality as follows,
\begin{align}\label{Z_lower_bound_new2}
Z(k,l)\geq&\frac{1}{\sqrt{w_1+w_2+w_3}}\int_{\mathcal{R}^2}\sqrt{w_1}S_1(k,l)+\sqrt{2w_2}S_2(k,l)\nonumber\\
&+\sqrt{w_3}S_3(k,l)d\mathbf{n}.
\end{align}

The optimal choice of $w_j,j=1,2,3$ is generally complicated and we aim to balance complexity and accuracy. Specifically, we set $w_3 = (\rho + \epsilon)^{-1}$ where $\epsilon$ prevents division by zero. The rationale is that when $\rho$ is small, the influence of IN and $S_3(k,l)$ is more significant, so the integral of $S_3(k,l)$ should receive a higher weight. Note that $S_1(k,l)$ and $S_2(k,l)$ can be neglected if $w_3 \gg w_1, w_2$. Conversely, if the IN is very weak and $\rho$ is close to zero, we maintain the weight of $S_3(k,l)$ based on the previous asymptotic analysis. By combining \eqref{Z_lower_bound_new1} and \eqref{Z_lower_bound_new2}, the lower bound for $Z(k,l)$ is given by
\begin{align}\label{Z_lower_bound_new}
Z(k,l)\geq&\max\Big(\int_{\mathcal{R}^2}S_3(k,l)d\mathbf{n},\int_{\mathcal{R}^2}\frac{1}{\sqrt{2+(\rho+\epsilon)^{-1}}}(S_1(k,l)\nonumber\\
&+S_2(k,l)+(\rho+\epsilon)^{-\frac{1}{2}}S_3(k,l))d\mathbf{n}\Big).
\end{align}

Next, the lower bound of each component in \eqref{Z_lower_bound_new} should be considered. The PLA method can be directly applied to establish the lower bounds based on the function family $\mathcal{T}_L([a,b],g,x)$ with minor modifications. However, a simpler bound can be obtained for the integral of $S_2(k,l)$. Specifically, $S_2(k,l)$ can be interpreted as the expectation over the tail section under a Gaussian distribution. Moreover, note that the function $(a + x)^{-b}$ with $x \ge 0$ and $b > 0$ is convex in $x$. Consequently, we have
\begin{align}\label{S_2_simple_lower_original}
&\int_{\mathcal{R}^2}S_2(k,l)d\mathbf{n}\nonumber\\
=&\sqrt{\frac{\rho(1-\rho)k_4}{k_3}}\mathbb{E}_{\mathbf{N}\sim\mathcal{N}(\mathbf{0},4\gamma_g^2\mathbf{I})}\Big(\Big(\frac{1+\Vert\mathbf{n}-\Delta\mathbf{s}(k,l)\Vert^2}{2\alpha\gamma_s^2}\Big)^{-\frac{\alpha+2}{4}}\Big)\nonumber\\
\geq&\sqrt{\frac{\rho(1-\rho)k_4}{k_3}}\Big(\frac{1+\mathbb{E}_{\mathbf{N}\sim\mathcal{N}(\mathbf{0},4\gamma_g^2\mathbf{I})}(\Vert\mathbf{n}-\Delta\mathbf{s}(k,l)\Vert^2)}{2\alpha\gamma_s^2}\Big)^{-\frac{\alpha+2}{4}}\nonumber\\
\triangleq&L(S_2(k,l)).
\end{align}

Based on the polar coordinate transformation, 
\begin{align}\label{S_2_simple_lower_med}
&\mathbb{E}_{\mathbf{N}\sim\mathcal{N}(\mathbf{0},4\gamma_g^2\mathbf{I})}(\Vert\mathbf{n}-\Delta\mathbf{s}(k,l)\Vert^2)\nonumber\\
=&\frac{1}{8\pi\gamma_g^2}\int_{0}^{+\infty}\int_{0}^{2\pi}r\exp\Big(-\frac{r^2}{8\gamma_g^2}\Big)(r^2-2r\Vert\Delta\mathbf{s}(k,l)\Vert\cos\phi\nonumber\\
&+\Vert\Delta\mathbf{s}(k,l)\Vert^2)d\phi dr\nonumber\\
=&8\gamma_g^2+\Vert\Delta\mathbf{s}(k,l)\Vert^2.
\end{align}

Substituting \eqref{S_2_simple_lower_med} into \eqref{S_2_simple_lower_original}, we have
\begin{align}\label{S_2_simple_lower_bound}
L(S_2(k,l))=8\pi\gamma_g^2\sqrt{\rho(1-\rho)k_3k_4}\Big(1+\frac{8\gamma_g^2+\Vert\Delta\mathbf{s}(k,l)\Vert^2}{2\alpha\gamma_s^2}\Big).
\end{align}

It can be observed that $L(S_2(k,l))$ is much simpler than the bounds obtained via the PLA method, although its performance may degrade in the low signal-to-noise ratio (SNR) region. This is because the slope of the Gaussian PDF in the mainlobe changes much more rapidly than in the tail region. Fixing the channel noise parameters, $\Vert \Delta \mathbf{s}(k,l) \Vert$ becomes large with the high SNR. In this case, the mainlobe of $(1 + \Vert \mathbf{n} - \Delta \mathbf{s}(k,l) \Vert^2)^{-(\alpha+2)/4}$ lies in the tail region of the PDF and the error introduced by Jensen's inequality is very small, asymptotically approaching zero. In other words, $L(S_2(k,l))$ converges to the exact value as the SNR increases.

For $S_3(k,l)$, the above approach is not applicable due to the square root operation in \eqref{Z_upper_bound}. The tail section is not integrable as its decay rate is insufficient. Therefore, the lower bound $L(S_3(k,l))$ must still rely on the PLA method, and the details are omitted for brevity.

\section{Constellation shaping}\label{section_constellation_shaping}
\subsection{Objective function transform}
Recall that constellation points and their associated probabilities are denoted by $\mathbf{s}_j$ and $p_j$ with $j=1,\cdots,M$, respectively. Consider the second-order power constraint on the input signals, with the maximum allowable power represented by $P_0$. Our objective is to maximize the cutoff rate via its lower bound. Accordingly, the optimization problem can be formulated as follows,
\begin{subequations}
\begin{align}
\mathbf{(P1):}\hspace{0.5cm}\max_{\mathbf{s}_1^M,p_1^M}\hspace{0.2cm}&L_{\text{CR}}(\mathbf{s}_1^M,p_1^M),\label{OP_1_a}\\
\text{s.t. }&\sum_{j=1}^{M}p_j=1,\label{OP_1_b}\\
&0\leq p_j\leq1,j=1,\cdots,M,\label{OP_1_c}\\
&\sum_{j=1}^{M}p_j\Vert\mathbf{s}_j\Vert^2\leq P_0.\label{OP_1_d}
\end{align}
\end{subequations}

The expression of $L_{\text{CR}}$ can be obtained in closed form from \eqref{CR_original}, \eqref{Z_upper_bound} and the derivations in Section~\ref{lower_bound_for_CR}. However, the constellation point $\mathbf{s}_j$ remains coupled inside special functions, which poses both computational and analytical challenges. To address this issue, we attempt to transform $U(S_3(k,l))$ into a simpler form.

It should be noted that the simplification strategy used for $S_2(k,l)$ cannot be applied to $S_3(k,l)$ since both the numerator and denominator of the argument in the Gaussian hypergeometric function depend on $n_1$. Consequently, $\Vert\mathbf{s}_j\Vert$ cannot be completely eliminated. A straightforward approach would be to expand $\prescript{}{2}F_1(\cdot)$ into a truncated series. However, this method suffers from two limitations. First, the series expansion is valid only when the magnitude of the argument is smaller than one. Otherwise, analytic continuation must be employed, which adds significant complexity as each $\prescript{}{2}F_1(\cdot)$ requires verification. Moreover, when the argument is close to one, the convergence of the series becomes slow, necessitating a large number of terms. This results in an objective function expressed as a summation of polynomials and rational functions. Although this resembles fractional programming, the convexity conditions for the numerator and denominator are not satisfied and theoretical convergence cannot be guaranteed \cite{paper28}. Furthermore, the high-order polynomial terms induced by truncation lead to numerical instabilities during the optimization process, such as the Runge phenomenon.

To circumvent these difficulties, we propose applying the PLA method at an earlier stage. For example, \eqref{S_3_med_1} is obtained by first integrating over $n_2$ and then applying PLA to simplify $J_1(x)$ and $J_3(x)$. In contrast, here we directly apply PLA with respect to $n_2$, namely,
\begin{align}\label{S_3_simple_bound}
&\int_{\mathcal{R}^2}S_3(k,l)d\mathbf{n}\nonumber\\
\leq&\int_{-\infty}^{+\infty}\sum_{h=1}^{K}\int_{r_{h-1}}^{r_h}\mathcal{T}_U([r_{h-1},r_h],J_4(n_1-\Vert\Delta\mathbf{s}(k,l)\Vert,n_2),n_2)\nonumber\\
&\times\mathcal{T}_U([r_{h-1},r_h],J_4(n_1,n_2),n_2)dn_2dn_1,
\end{align}
where $J_4(x,y)\triangleq(2\alpha\gamma_s^2+x^2+y^2)^{-\frac{\alpha+2}{4}}$. After evaluating \eqref{S_3_simple_bound}, the remaining integral over $n_1$ can be derived in closed form using \eqref{identity_2}. It can be verified that the new upper bound of $\int_{\mathcal{R}^2} S_3(k,l) d\mathbf{n}$ takes the linear form $\tau_1 \Vert \Delta \mathbf{s}(k,l) \Vert + \tau_0$. The explicit expressions for $\tau_1$ and $\tau_0$ are obtainable but omitted for brevity. Similarly, the lower bound of $\int_{\mathcal{R}^2} S_2(k,l) d\mathbf{n}$ is also linear in $\Vert \Delta \mathbf{s}(k,l) \Vert$. Consequently, the objective function can be written as
\begin{align}\label{new_objevtive_function}
&L_{\text{CR}}(\mathbf{s}_{1,t}^M,p_{1,t}^M)\nonumber\\
=&-\log\bigg(\sum_{j=1}^{M}p_{j,t}^2+\underbrace{\sum_{j=1}^{M-1}\sum_{k=j+1}^{M}p_{j,t}p_{k,t}\tilde{Z}_t(k,l)}_{\triangleq\bar{Z}_t(\mathbf{s}_{1,t}^M,p_{1,t}^M)}\bigg),
\end{align}
where
\begin{align}
\tilde{Z}_t(k,l)\triangleq&4\rho k_3\pi\gamma_g^2\exp\Big(-\frac{\Vert\Delta\mathbf{s}_t(k,l)\Vert^2}{16\gamma_g^2}\Big)\nonumber\\
&+\varrho_1\Vert\Delta\mathbf{s}_t(k,l)\Vert+\varrho_0,
\end{align}
where the subscript $t$ denotes the values at the $t$-th iteration. Here, $\mathbf{s}_{1,t}^M \triangleq (\mathbf{s}_{1,t}\cdots, \mathbf{s}_{M,t})$ represents the array of signal vectors. The constants $\varrho_1$ and $\varrho_0$ can be computed following the procedures used to obtain the upper bound of the CR.

\subsection{Iteration algorithm}
Based on \eqref{new_objevtive_function}, the primal problem $(\mathbf{P1})$ can be decomposed into two decoupled subproblems, namely probabilistic shaping and geometric shaping. For example,
\begin{subequations}
\begin{align}
\mathbf{(P2A)}:\hspace{0.5cm}\min_{p_{1,t}^M}\hspace{0.2cm}&\sum_{j=1}^{M}p_{j,t}^2+\bar{Z}_t(\mathbf{s}_{1,t}^M,p_{1,t}^M),\label{OP_2_a}\\
\text{s.t. }&\sum_{j=1}^{M}p_{j,t}=1, \nonumber\\
&0\leq p_{j,t}\leq1,j=1,\cdots,M, \nonumber\\
\mathbf{(P2B)}:\hspace{0.5cm}\min_{\mathbf{s}_{1,t}^M}\hspace{0.2cm}&\bar{Z}_t(\mathbf{s}_{1,t}^M,p_{1,t}^M),\label{OP_2_d}\\
\text{s.t. }&\sum_{j=1}^{M}p_{j,t}\Vert\mathbf{s}_{j,t}\Vert^2\leq P_0.\nonumber
\end{align}
\end{subequations}

The problem $\mathbf{(P2A)}$ is a standard quadratic programming (QP) and convex in terms of $p_{j,t}$. The optimal solution can be derived using the Lagrange multiplier method. For example,
\begin{align}\label{optimal_solution_p_j}
p_{j,t}^*=\frac{1}{M}\sum_{k=1}^{M}\sum_{l=1}^{M}p_{j,t}\tilde{Z}_t(k,l)-\sum_{k=1,k\neq j}^{M}p_{k,t}\tilde{Z}_t(j,k).
\end{align}

With the aid of $0\leq\tilde{Z}_t(k,l)\leq1$, it can be shown that $p_{j,t}^*$ automatically satisfies $p_{j,t}^*\leq 1$. To enforce non-negativity, the projection method is applied, i.e., $p_{1,t+1}^M=\text{Proj}_{\Omega_+}((p_{1,t}^M)^*)$ where $\Omega_+\triangleq\{p_1^M:0\leq p_{j}\leq1,j=1,\cdots,M,\sum_{j=1}^{M}p_{j}=1\}$. The target set of this projection is the standard probabilistic simplex and the detailed projection procedures are omitted for simplicity.

In the next, we mainly focus on the geometric distribution updating. There are no solutions in closed form and we combine the gradient descent and projection for iteration, i.e.,
\begin{align}\label{iteration_s_j}
\mathbf{s}_{j,t+1}=&\text{Proj}_{\Omega_P(j)}(\mathbf{s}_{j,t}-\mu \nabla_{\mathbf{s}_{j,t}}\bar{Z}_t(p_{1,t+1}^M,\mathbf{s}_{1,t+1}^{j-1},\mathbf{s}_{j,t}^{M}))\nonumber\\
\triangleq&\text{Proj}_{\Omega_P(j)}(\hat{\mathbf{s}}_{j,t}),
\end{align}
where $\Omega_P(j)\triangleq\{\mathbf{s}:\sum_{k=1}^{j-1}p_{k,t+1}\Vert\mathbf{s}_{k,t+1}\Vert^2+\sum_{k=j+1}^{M}p_{k,t}\Vert\mathbf{s}_{k,t}\Vert^2+p_{j,t+1}\Vert\mathbf{s}\Vert^2\leq P_0\}$. The projected point can also be computed using the Lagrange multiplier method as follows,
\begin{align}\label{projection_s_j}
\mathbf{s}_{j,t+1}=\frac{\hat{\mathbf{s}}_{j,t}}{\sqrt{p_{j,t+1}}\Vert\hat{\mathbf{s}}_{j,t}\Vert}\sqrt{P_0-P_{j,t}},
\end{align}
where
\begin{align}
P_{j,t}\triangleq\sum_{k=1}^{j-1}p_{k,t+1}\Vert\mathbf{s}_{k,t+1}\Vert^2+\sum_{k=j+1}^{M}p_{k,t+1}\Vert\mathbf{s}_{k,t}\Vert^2.
\end{align}

The selection of the iteration step $\mu$ in \eqref{iteration_s_j} is discussed in the next subsection. The overall constellation shaping algorithm is summarized in Algorithm~\ref{algorithm_2}.

\begin{algorithm}[!ht]
\renewcommand{\algorithmicrequire}{\textbf{Input:}}
\renewcommand{\algorithmicensure}{\textbf{Output:}}
\caption{Constellation geometric and probabilistic shaping under the mixed channel noise.}\label{algorithm_2}
\begin{algorithmic}[1]
\Require Original constellation points $\mathbf{s}_{1,0}^{M}$ and corresponding probabilities $p_{1,0}^{M}$. 
\Ensure Optimized constellation points $\mathbf{s}_{1,0}^{M}$ and corresponding probabilities $p_{1,0}^{M}$. 
\State $e=+\infty$, $t=0$, $I_{\text{max}}=500$.
\While {$t\leq I_{\text{max}}$ and $e\geq\epsilon=10^{-4}$}
\State Obtain $p_{j,t+1}$ via \eqref{optimal_solution_p_j} and probabilistic simplex projection.
\State Obtain $\mathbf{s}_{j,t+1}$ via \eqref{iteration_s_j} and \eqref{projection_s_j}.
\State $e_j=\max(|p_{j,t+1}-p_{j,t}|,\Vert\mathbf{s}_{j,t+1}-\mathbf{s}_{j,t}\Vert),j=1,\cdots,M$.
\State $e=\max(e_1,\cdots,e_M)$.
\State $t=t+1$.
\EndWhile
\end{algorithmic}
\end{algorithm}

\textbf{Remark 1:} The computational complexity of determining $\varrho_1$ and $\varrho_0$ is minor and depends solely on the division resolution of the PLA method. Compared to the resolution required for numerical integration, the PLA resolution can be much lower and therefore, the computational burden is acceptable.

\textbf{Remark 2:} Algorithm~\ref{algorithm_2} is a gradient-based method. More efficient alternatives, such as Newton's method or the Adam optimizer, could potentially accelerate convergence. \textbf{Our main contribution herein lies in transforming the original problem into a more tractable form and providing the associated theoretical analysis presented in the next subsection.}

\textbf{Remark 3:} The preceding analysis is conducted at the constellation point level. In practical implementation, the source bit distribution may not be uniform. To map a bit sequence to modulated constellation points with an arbitrary distribution, a powerful tool is the constant composition distribution matcher (CCDM) \cite{paper29}. This method can be directly applied to address the mapping problem. Hence, the implementation details are omitted.

\subsection{Convergence analysis}
Before proving the convergence of the Algorithm \ref{algorithm_2}, the following theorem is first provided.
\begin{theorem}\label{theorem_1}
\textit{Denote $Z(k,l)$ in \eqref{CR_original} equivalently as $Z(\Delta \mathbf{s}(k,l))$. Define $\kappa(\Vert \Delta \mathbf{s}(k,l)\Vert)$ such that
\begin{align}
\nabla_{\mathbf{s}_{k}}Z(\Delta\mathbf{s}(k,l))=\kappa(\Vert\Delta\mathbf{s}(k,l)\Vert)\Delta\mathbf{s}(k,l).
\end{align}}

\textit{Then, $Z(k,l)$ is $L(k,l)$-smooth with respect to $\mathbf{s}_k$ and $\mathbf{s}_l$ where
\begin{align}\label{Lipschitz_constant_for_k_l}
L(k,l)=&\sup\limits_{\mathbf{s}_h\in\Omega_P(h),\mathbf{s}_l\in\Omega_P(l)}|\nabla_{\Vert\Delta\mathbf{s}(h,l)\Vert}\kappa(\Vert\Delta\mathbf{s}(h,l)\Vert)|\nonumber\\
&\times\Vert\Delta\mathbf{s}(h,l)\Vert+\max(Z(\mathbf{0}),|\nabla\kappa(\mathbf{0})|),
\end{align}
where $\nabla\kappa(\mathbf{0})$ herein represents the value of the derivative of $\kappa(\cdot)$ at the origin.}
\end{theorem}
\begin{IEEEproof}
The proof is relegated to Appendix \ref{appendix_1}.
\end{IEEEproof}

\begin{corollary}\label{corollary_1}
\textit{$\bar{Z}$ is $L$-smooth with respect to $\mathbf{s}_j,j=1,\cdots,M$ and
\begin{align}
L=\sum_{k=1}^{M-1}\sum_{l=k+1}^{M}p_kp_lL(k,l),
\end{align}
where $L(k,l)$ is given in \eqref{Lipschitz_constant_for_k_l}.}
\end{corollary}

Note that Theorem \ref{theorem_1} applies to $Z(k,l)$, which differs from $\bar{Z}_t$ in the optimization problems $(\mathbf{P2A})$ and $(\mathbf{P2B})$. Nevertheless, $\bar{Z}_t$ satisfies almost all the assumptions of Theorem \ref{theorem_1}. First, $\bar{Z}_t$ asymptotically converges to $\bar{Z}$. Therefore, $V_t(h,l)$ vanishes as $\Vert \Delta \mathbf{s}_t(h,l) \Vert \to +\infty$, where the subscript $t$ distinguishes parameters associated with $\bar{Z}$ and $\bar{Z}_t$. Hence, $\sup\limits_{\mathbf{s}_{h,t} \in \Omega_P(h), \mathbf{s}_{l,t} \in \Omega_P(l)} V_t(h,l)$ is finite. Second, $\bar{Z}_t$ is obviously finite. Although its convexity cannot be formally proved, the step $(a)$ in \eqref{decompose_Lipschitz} still holds with minor modifications, i.e.,
\begin{align}\label{reality_lipschitz}
&|\kappa_t(h,l)-\kappa_t(k,l)|\nonumber\\
\leq&\sup\limits_{\mathbf{s}_{h,t}\in\Omega_P(h),\mathbf{s}_{l,t}\in\Omega_P(l)}|\nabla_{\Vert\Delta\mathbf{s}(h,l)\Vert}\kappa_t(\Vert\Delta\mathbf{s}(h,l)\Vert)|\Vert\mathbf{s}_{k,t}-\mathbf{s}_{h,t}\Vert.
\end{align}

Similarly, since the asymptotic behavior of $\bar{Z}_t$ matches that of $\bar{Z}$, the difference in the $\sup$ in \eqref{reality_lipschitz} occurs only in the finite low SNR region. Consequently, the RHS of \eqref{reality_lipschitz} is finite. Based on this analysis, the following corollary follows directly.
\begin{corollary}\label{corollary_2}
\textit{$\bar{Z}_t$ is $L$-smooth with respect to $\mathbf{s}_j,j=1,\cdots,M$ where
\begin{align}
\bar{L}=&\sum_{k=1}^{M-1}\sum_{l=k+1}^{M}p_kp_l(\max(\tilde{Z}(\mathbf{0}),|\nabla\kappa(\mathbf{0})|)\nonumber\\
&+\sup\limits_{\mathbf{s}_{h,t}\in\Omega_P(h),\mathbf{s}_{l,t}\in\Omega_P(l)}|\kappa(\Vert\Delta\mathbf{s}(h,l)\Vert)|\Vert\Delta\mathbf{s}(h,l)\Vert).
\end{align}}
\end{corollary}

With Corollary \ref{corollary_2}, the convergence of Algorithm \ref{algorithm_2} can be guaranteed by setting $\mu\leq2\bar{L}^{-1}$. Then, we analyze convergent behaviors of constellation point sequence and the corresponding function value sequence. Without loss of generality, we focus on the $j$-th point and examine the convergence behavior of ${\mathbf{s}_{j,t}}$ and ${\bar{Z}(\Delta\mathbf{s}_t(k,l))}$ with respect to $t$. Since Algorithm \ref{algorithm_2} is a standard PGD approach, existing convergence analysis techniques can be applied. Accordingly, we present the following two theorems for ${\mathbf{s}_{j,t}}$ and ${\bar{Z}(\Delta\mathbf{s}_t(k,l))}$, providing only a proof sketch for brevity.

\begin{theorem}\label{theorem_2}
\textit{Assume that sequences $\{L_{\text{CR}}(\mathbf{s}_{1,t}^M,p_{1,t}^M)\}$ and $\{\bar{Z}_t(\mathbf{s}_{1,t}^M,p_{1,t}^M)\}$, $t=0,1,\cdots$ are generated by Algorithm \ref{algorithm_2}. Define $\mathcal{P}(\mu,x)=\mu^{-1}(x-\text{Proj}_{\Omega_{P}(j)}(x-\mu\nabla\bar{Z}_t))$ and then, we have
\begin{enumerate}
\item{The sequence $\{L_{\text{CR}}(\mathbf{s}_{1,t}^M,p_{1,t}^M)\}$ is bounded and monotonically decreasing, and therefore converges to $L_{\text{CR}}^*$.}
\item{If we fix $p_{k,t},k=1,\cdots,M$ and $\mathbf{s}_{k,t},k=1,\cdots,j-1,j+1,\cdots,M$, and only consider the iteration of $\mathbf{s}_{j,t}$, then we have
    \begin{equation}\label{second_claim_1}
    \bar{Z}_t(\mathbf{s}_{j,t+1})\leq \bar{Z}_t(\mathbf{s}_{j,t})-\frac{\mu}{2}\Vert\mathcal{P}(\mu,\mathbf{s}_{j,t})\Vert^2,
    \end{equation}
    \begin{equation}\label{second_claim_2}
    \min\limits_{0\leq t\leq T}\Vert\mathcal{P}(\mu,\mathbf{s}_{j,t})\Vert^2\leq\frac{2(\bar{Z}_t(\mathbf{s}_{j,0})-\bar{Z}_t^*)}{\mu T},
    \end{equation}
    }
\end{enumerate}
where $\bar{Z}_t(\mathbf{s}_{j,t+1})$ denotes the value of $\bar{Z}_t(\cdot)$ evaluated at $\mathbf{s}_{j,t+1}$, with all other arguments fixed within the feasible set.}
\end{theorem}
\hspace{2em}\textit{Proof sketch:} For the first claim, we need to show that the function value decreases after each iteration. When updating $p_{j,t}$, the optimal $p_{j,t}^*$ is obtained and the objective function is quadratic. Hence, the function value decreases even if the projection is active. As for $\mathbf{s}_{j,t}$, based on Theorem \ref{theorem_1}, setting the update step smaller than the reciprocal of the Lipschitz constant guarantees monotonicity.

According to the definition of $\mathcal{P}(\mu,x)$, \eqref{iteration_s_j} can be written as $\mathbf{s}_{j,t+1} = \mathbf{s}_{j,t} - \mu \mathcal{P}(\mu, \mathbf{s}_{j,t})$. Combining this with the second-order upper bound ensures that the function value decreases after each iteration, leading to \eqref{second_claim_1}. Regarding \eqref{second_claim_2}, it follows by summing \eqref{second_claim_1} from the first to the $T$-th iteration and taking the average. Moreover, \eqref{second_claim_2} implies a convergence rate of $\mathcal{O}(1/\sqrt{T})$. $\hfill\blacksquare$

\textbf{Remark 4:} The second claim may appear inconsistent with the optimization procedure in Algorithm \ref{algorithm_2}. In fact, it provides a stronger result because in the complete sequence, the function value after updating $\mathbf{s}_{j,t}$ is not the original value at the next epoch when updating $\mathbf{s}_{j,t+1}$. Due to monotonicity, if we extract the subsequence $\{L_{\text{CR}}(p_{1,t+1}^M, \mathbf{s}_{1,t}^j, \mathbf{s}_{j+1,t}^M)\}$, it is always upper bounded by the sequence in Theorem \ref{theorem_2}.

\begin{theorem}\label{theorem_3}
\textit{Assume that the sequence $\{\mathbf{s}_{j,t}\},t=0,1,\cdots$ is generated by Algorithm \ref{algorithm_2}. Let $\{\tilde{\mathbf{s}}_{j,t}\}$ be a subsequence of $\{\mathbf{s}_{j,t}\}$ that converges to $\tilde{\mathbf{s}}_j^*$. Then, the limit point satisfies $0\in\nabla_{\mathbf{s}_{j}} L_{\text{CR}}(\tilde{\mathbf{s}}_{j}^*)+N_c(\tilde{\mathbf{s}}_{j}^*)$ where $N_c(x)$ denotes the normal cone of the feasible set at the point $x$.}
\end{theorem}
\hspace{2em}\textit{Proof sketch:} According to Theorem \ref{theorem_2}, the sequence of function values converges. Hence, there exists a subsequence ${\tilde{\mathbf{s}}_{j,t}}$ such that $\tilde{\mathbf{s}}_{j,t}\rightarrow \tilde{\mathbf{s}}_j^*$ and $\Vert \mathcal{P}(\mu, \tilde{\mathbf{s}}_{j,t}) \Vert \rightarrow 0$. Then, by applying the first-order optimality condition of the projection operation and performing some manipulations, the theorem is established. Moreover, all cluster points of the sequence satisfy the Karush-Kuhn-Tucker (KKT) conditions. $\hfill\blacksquare$

\textbf{Remark 5:} From the derivation of $\bar{Z}_t$, it is clear that $\bar{Z}_t$ is composed of integrals of piecewise linear functions and rational terms. Since these integrals are always finite, the Kurdyka–Łojasiewicz (KL) condition is satisfied \cite{book2}. Consequently, Theorem \ref{theorem_3} can be strengthened: the sequence ${\mathbf{s}_{j,t}}$ converges to a single stationary point rather than possibly multiple distinct points.

\section{Simulations}\label{section_simulations}
In this section, we evaluate the tightness of the derived bounds and the performance of the proposed constellation optimization algorithm. For the mixed noise scenario, the conventional SNR is not applicable due to the infinite second moment. Therefore, we adopt the generalized signal-to-noise ratio (GSNR), defined as
\begin{align}\label{GSNR_definition}
\text{GSNR(dB)}=10\log_{10}\frac{P_s}{2(\gamma_g^2+\gamma_s^2)},
\end{align}
where $P_s$ denotes the power of the transmitted signals. Since \eqref{baseband_PDF} involves four parameters, we focus on those that significantly impact the mixed noise. Accordingly, we fix $\gamma_s = \gamma_g = 1$ and vary $\alpha$ and $\rho$. In particular, we set $(\alpha, \rho)$ to $(1.2, 0.2)$ and $(1.8, 0.8)$ to represent scenarios with weak and strong impulsiveness, respectively.

\subsection{Bounds verifications}
\begin{figure*}[htbp]
\centering
\subfloat[$\alpha=1.2$, $\rho=0.2$, $S_2(k,l)$]{\includegraphics[width=1.75in]{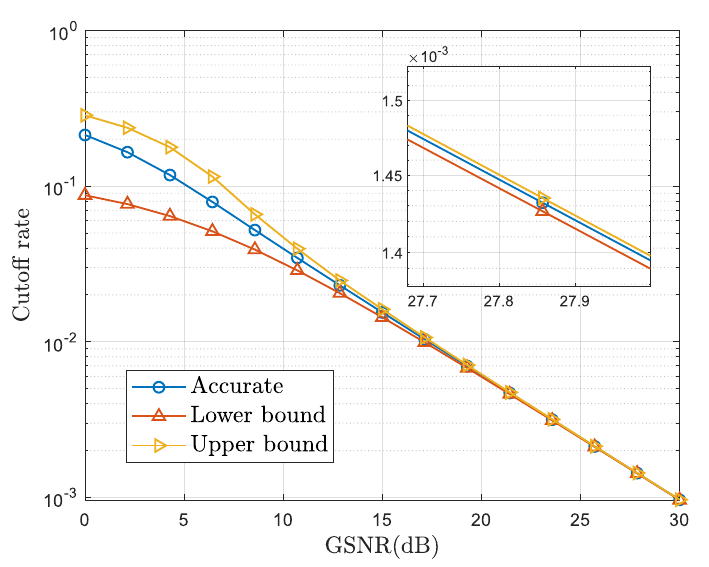}\label{fig_1_a}}
\subfloat[$\alpha=1.8$, $\rho=0.8$, $S_2(k,l)$]{\includegraphics[width=1.75in]{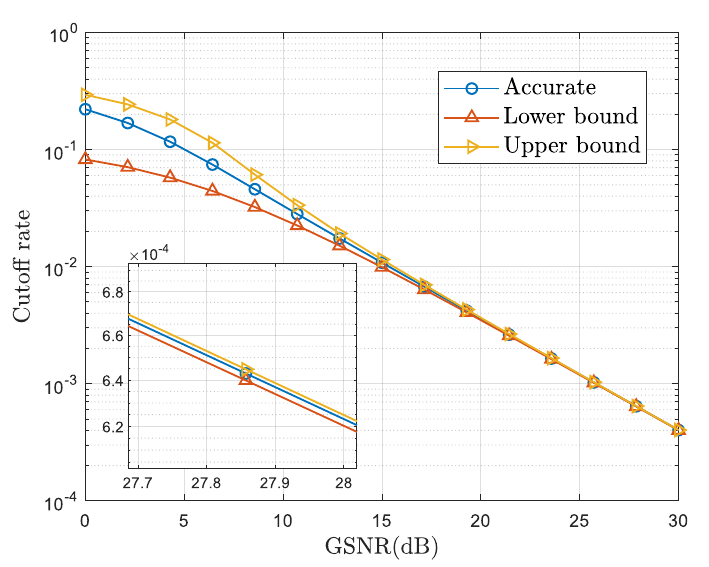}\label{fig_1_b}}
\subfloat[$\alpha=1.2$, $\rho=0.2$, $S_3(k,l)$]{\includegraphics[width=1.75in]{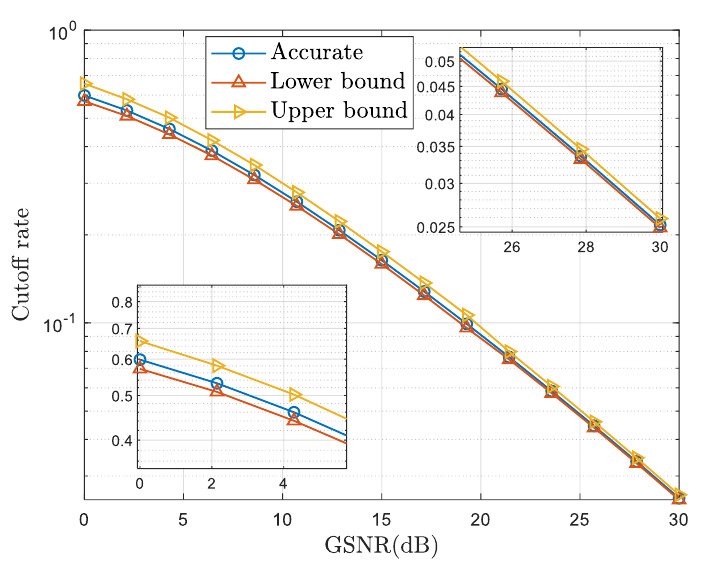}\label{fig_1_c}}
\subfloat[$\alpha=1.8$, $\rho=0.8$, $S_3(k,l)$]{\includegraphics[width=1.75in]{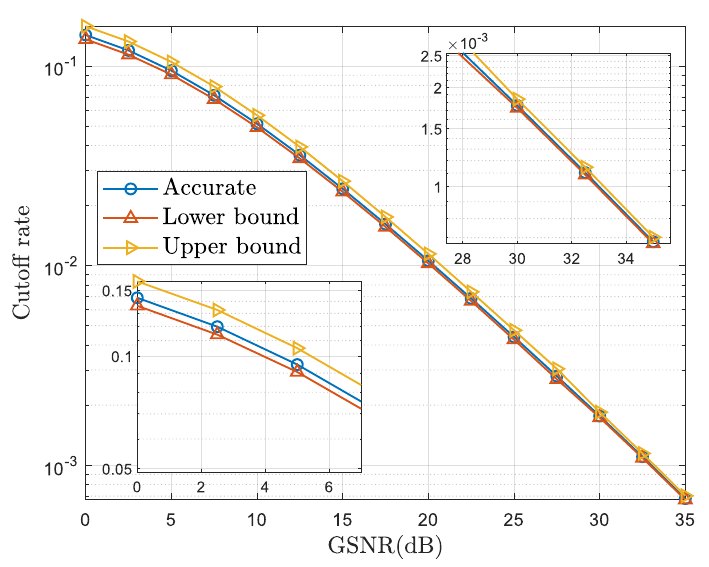}\label{fig_1_d}}
\caption{Lower and upper bounds for integrals of $S_2(k,l)$ and $S_3(k,l)$ under various $\alpha$ and $\rho$.}
\label{fig_1}
\end{figure*}

We first verify the tightness of the bounds for the integrals of $S_2(k,l)$ and $S_3(k,l)$, as they directly determine the final performance of the CR bounds. Without loss of generality, we set the underlying vector difference as $\Delta\mathbf{s}(k,l) = A(1 - 1j)$, where $A$ denotes its amplitude. The results are shown in Fig. \ref{fig_1}. In Figs. \ref{fig_1_a} and \ref{fig_1_b}, the upper bound corresponds to \eqref{S_2_upper_bound}, while the lower bound corresponds to \eqref{S_2_simple_lower_bound}. For Fig. \ref{fig_1_c} and \ref{fig_1_d}, both bounds are obtained using the PLA method.

From the results, it can be observed that all bounds exhibit asymptotic behavior. The numbers of divided intervals for the PLA of $S_2(k,l)$ and $S_3(k,l)$ are 45 and 60, respectively. Generally, $S_3(k,l)$ requires a higher resolution because the underlying functions are more complex and decay only at a polynomial rate. The PLA method with the above resolutions provides excellent performance, although a higher division resolution can further improve tightness. Furthermore, the convergence rate in Fig. \ref{fig_1_a} and \ref{fig_1_b} is significantly faster than that in Fig. \ref{fig_1_c} and \ref{fig_1_d}. This is because the exponential term in $S_2(k,l)$ decays rapidly as the GSNR increases, causing the approximation error to diminish quickly.

Next, we compare the bounds with the exact CR under the conventional QPSK scheme. Compared to the previous simulation for a single constellation pair, the main differences in the CR bounds arise from two factors. First, all integrals of $S_2(k,l)$ and $S_3(k,l)$ over different $(k,l)$ pairs must be summed. More importantly, the outer inequalities introduce additional approximation errors. Consequently, the tightness of the CR bounds is generally worse than that for a single constellation pair. However, this error can be partially mitigated due to the logarithmic operation in the CR expression.

\begin{figure}[htbp]
\centering
\subfloat[$\alpha=1.2$, $\rho=0.2$]{\includegraphics[width=1.75in]{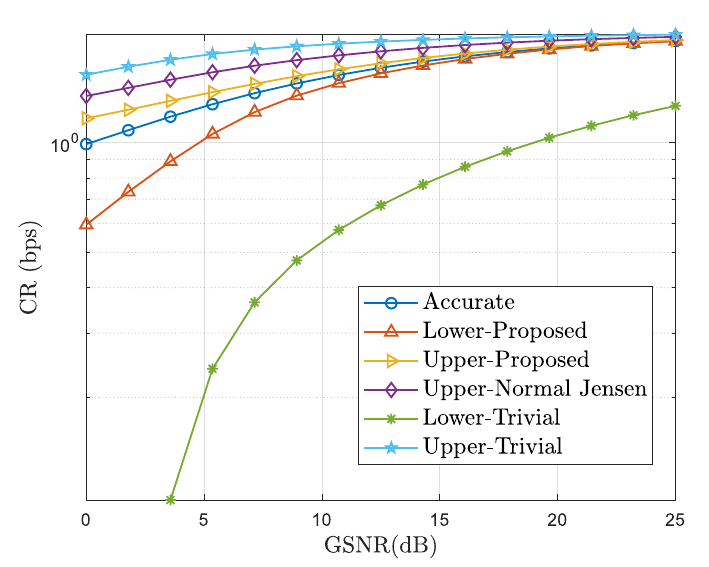}\label{fig_2_a}}
\subfloat[$\alpha=1.8$, $\rho=0.8$]{\includegraphics[width=1.75in]{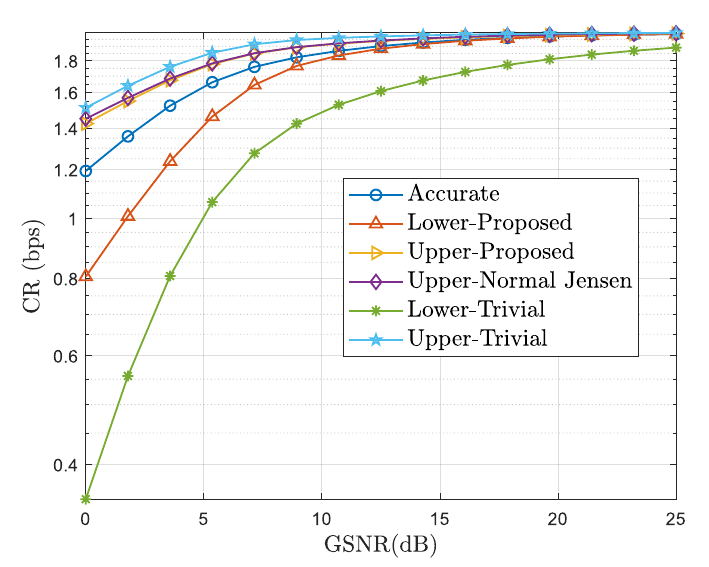}\label{fig_2_b}}
\caption{Lower and upper CR bounds for the QPSK modulation scheme under various $\alpha$ and $\rho$.}
\label{fig_2}
\end{figure}

The numerical comparisons are presented in Fig. \ref{fig_2}, where we also include bounds obtained using different outer inequalities. `Lower-Proposed' and `Upper-Proposed' denote the bounds based on the inner approximations shown in Fig. \ref{fig_1}. `Upper-Normal Jensen' corresponds to the upper bound using the outer approximation via standard Jensen's inequality. `Upper-Trivial' represents the bound derived from the triangle inequality applied to the integration of $S_2(k,l)$ and $S_3(k,l)$, with a similar notation used for `Lower-Trivial'. Overall, the proposed lower and upper bounds are both tight and asymptotic, and the gap mainly stems from the approximation error of the outer bounds. In particular, when $\alpha$ and $\rho$ are small, the lower bound derived from \eqref{Z_lower_bound_new} is significantly tighter than that obtained using only the standard Jensen's inequality, demonstrating the effectiveness of our outer bounds. As $\rho$ increases, the weight for $S_3(k,l)$ decreases, making it very similar to the normal Jensen's inequality and thus reducing the improvement.

\subsection{Cutoff rate comparison}
\begin{figure*}[htbp]
\centering
\subfloat[$\alpha=1.2$, $\rho=0.2$, $0$\scriptsize{dB}]{\includegraphics[width=1.75in]{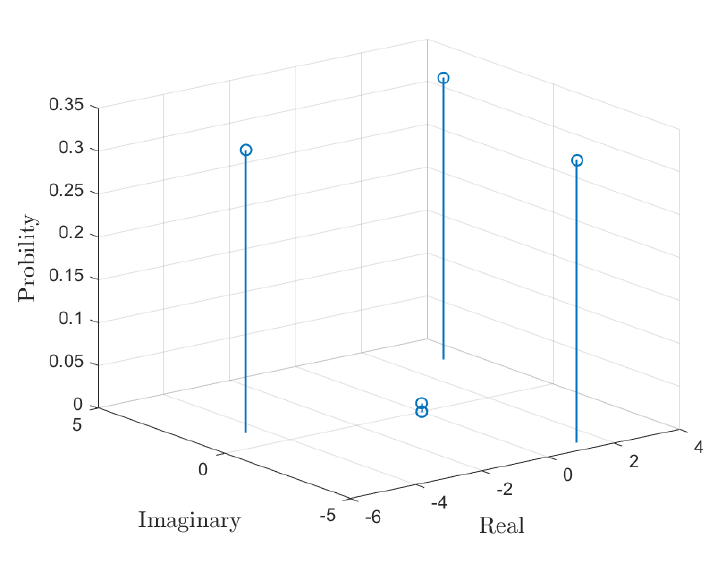}\label{fig_3_a}}
\subfloat[$\alpha=1.2$, $\rho=0.2$, $16$\scriptsize{dB}]{\includegraphics[width=1.75in]{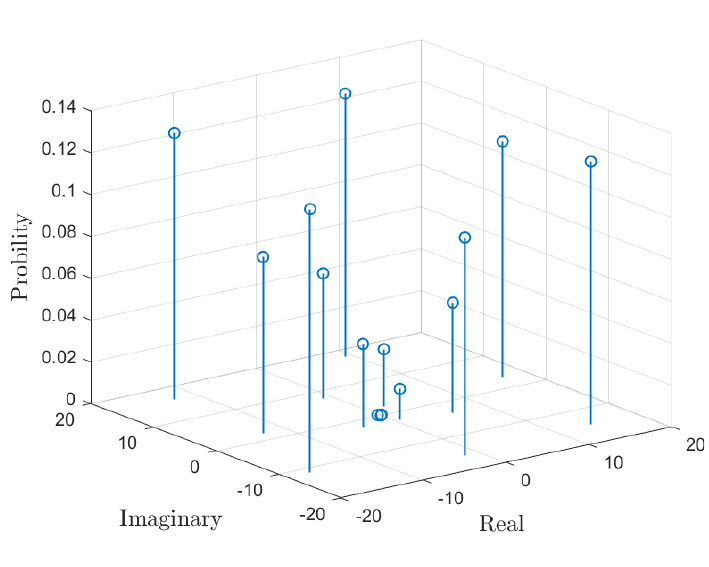}\label{fig_3_b}}
\subfloat[$\alpha=1.8$, $\rho=0.8$, $0$\scriptsize{dB}]{\includegraphics[width=1.75in]{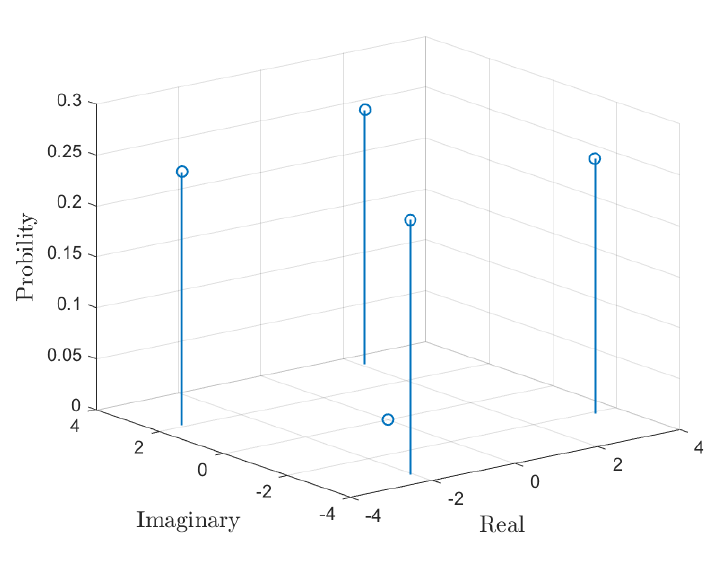}\label{fig_3_c}}
\subfloat[$\alpha=1.8$, $\rho=0.8$, $16$\scriptsize{dB}]{\includegraphics[width=1.75in]{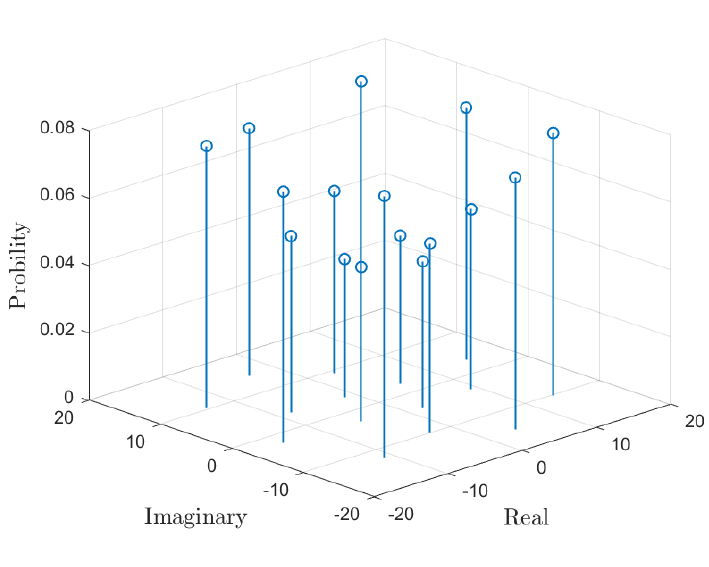}\label{fig_3_d}}
\caption{Geometric and probability distribution of the optimized constellation points under different GSNR, $\alpha$ and $\rho$.}
\label{fig_3}
\end{figure*}

In this subsection, we visualize the optimized constellation points and provide corresponding CR comparisons with the conventional constellation scheme. We set $M = 16$, which is the number of feasible constellation points. The optimized constellations under different GSNR, $\alpha$ and $\rho$ are shown in Fig. \ref{fig_3}.

From Fig. \ref{fig_3_a} and \ref{fig_3_c}, it can be observed that many points are inactive. This is reasonable because in the low GSNR region, the signal power is limited. To maximize the CR, power must be used efficiently, which corresponds to allocating it to a small subset of points. Although this reduces the total entropy of the transmitted signal, the minimum distance between active points increases, resulting in lower error probability and higher CR compared to the conventional scheme. As the GSNR increases, more points become active and their geometric positions are optimized, as illustrated in Fig. \ref{fig_3_b} and \ref{fig_3_d}. Regarding different $\alpha$ and $\rho$, it can be observed that fewer constellation points are active for smaller $\alpha$ and $\rho$ at a fixed GSNR. This is consistent with expectations since stronger impulsiveness can rapidly increase detection errors when more points are used. Therefore, limiting the number of active points is a prudent strategy to avoid significant CR loss.

Next, we compare the CR performance between the optimized constellations and several baseline schemes. Here, we set $M = 16$ and consider four baselines: `Conventional', `Only geo', `Only pro' and `WGNC'. `Conventional' refers to the standard square 16-QAM constellation. `Only geo' and `Only pro' correspond to constellations shaped solely by geometric and probabilistic methods, respectively. `WGNC' denotes the optimized constellation for WGN scenarios without considering the IN component.

\begin{figure}[htbp]
\centering
\subfloat[$\alpha=1.2$, $\rho=0.2$]{\includegraphics[width=1.75in]{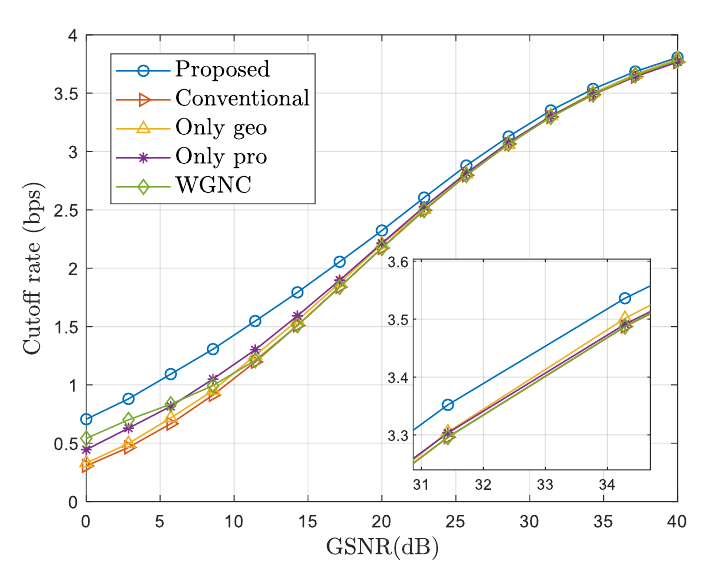}\label{fig_4_a}}
\subfloat[$\alpha=1.8$, $\rho=0.8$]{\includegraphics[width=1.75in]{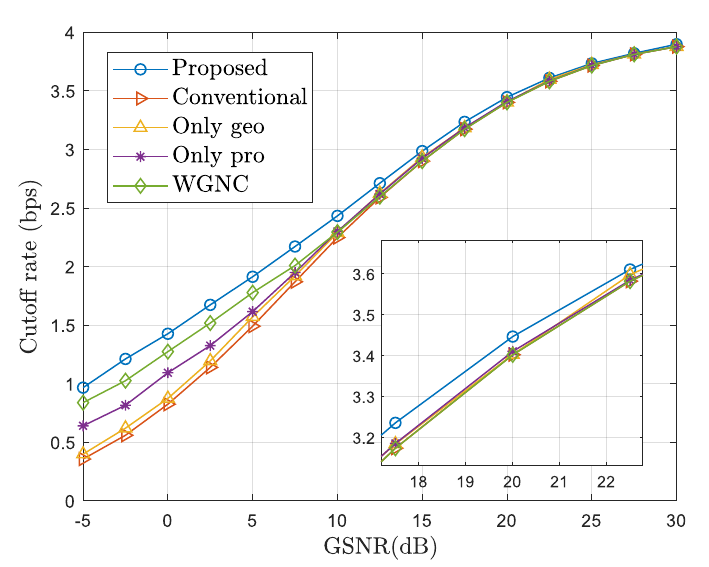}\label{fig_4_b}}
\caption{CR comparison between optimized constellation and baselines under various $\alpha$ and $\rho$.}
\label{fig_4}
\end{figure}

The simulation results are presented in Fig. \ref{fig_4}. The proposed optimization scheme achieves the highest CR across all GSNRs and noise parameter settings, particularly in the low and medium GSNR regimes. Specifically, improvements of at least 1.9 dB and 0.4 dB are observed when the CR reaches 2 bps and 3.5 bps, respectively. At high GSNR, the CR improvement becomes less pronounced due to the modulation order constraint, which is a common phenomenon in constellation optimization observed in other scenarios \cite{paper21,paper31,paper32}. Moreover, the CR gain of `Only geo' exceeds that of `Only pro', indicating that geometric shaping has a more dominant impact on CR. It has been shown that the performance gains achieved by probabilistic shaping can also be attained through appropriate geometric shaping \cite{paper30}, implying that the gains from probabilistic shaping can be viewed as a `subset' of those from geometric shaping. Finally, a notable CR degradation is observed for `WGNC' compared to the proposed scheme, confirming that the effect of IN cannot be neglected, especially under strong impulsiveness.

\section{Conclusions}\label{section_conclusions}
In this paper, we investigated rate bounds and constellation shaping under non-Gaussian mixed noise. First, we derived the baseband noise model for the mixed noise. Using this model, we designed lower and upper bounds for the cutoff rate based on the PLA approach. Furthermore, the lower bound of the CR was employed to optimize both the geometric and probabilistic distributions of the constellation points. PGD was applied for the optimization and its convergence was analyzed. Numerical results demonstrated that the proposed bounds were tight and exhibited asymptotic behavior. Additionally, the optimized constellations achieved higher CR, particularly in the low and medium GSNR regimes.

\begin{appendices}
\section{Proof of Theorem \ref{theorem_1}}\label{appendix_1}
\setcounter{equation}{0}
\renewcommand{\theequation}{A.\arabic{equation}}
We first rewrite $Z(k,l)$ as a function of $\mathbf{s}_k$ with $\mathbf{s}_l$ held fixed, i.e., $Z(\mathbf{s}_k)$. The time index is omitted since our goal is to determine the global Lipschitz constant. For any fixed $\mathbf{s}_l$, it can be verified that $\nabla{\mathbf{s}_k} Z(\mathbf{s}_k)$ takes the form $\kappa(\Vert \Delta \mathbf{s}(k,l) \Vert)(\mathbf{s}_k - \mathbf{s}_l)$ where $\kappa(\Vert \Delta \mathbf{s}(k,l) \Vert)$ is a finite function dependent on the chosen $\Vert \Delta \mathbf{s}(k,l) \Vert$. Then,
\begin{align}
&\Vert\nabla_{\mathbf{s}_k}Z(\mathbf{s}_k)-\nabla_{\mathbf{s}_k}Z(\mathbf{s}_h)\Vert\nonumber\\
=&\Vert\kappa(\Vert\Delta\mathbf{s}(k,l)\Vert)\Delta\mathbf{s}(k,l)-\kappa(\Vert\Delta\mathbf{s}(h,l)\Vert)\Delta\mathbf{s}(h,l)\Vert.
\end{align}

First, we assume that $\kappa(\Vert\Delta\mathbf{s}(h,l)\Vert)\geq\kappa(\Vert\Delta\mathbf{s}(k,l)\Vert)$ and then,
\begin{align}\label{decompose_Lipschitz}
&\Vert\kappa(\Vert\Delta\mathbf{s}(k,l)\Vert)\Delta\mathbf{s}(k,l)-\kappa(\Vert\Delta\mathbf{s}(h,l)\Vert)\Delta\mathbf{s}(h,l)\Vert\nonumber\\
\overset{(a)}{\leq}&|\kappa(\Vert\Delta\mathbf{s}(h,l)\Vert)-\kappa(\Vert\Delta\mathbf{s}(k,l)\Vert)|\Vert\Delta\mathbf{s}(h,l)\Vert\nonumber\\
&+|\kappa(\Vert\Delta\mathbf{s}(k,l)\Vert)|\Vert\Delta\mathbf{s}(k,h)\Vert\nonumber\\
\overset{(b)}{\leq}&|\nabla_{\Vert\Delta\mathbf{s}(h,l)\Vert}\kappa(\Vert\Delta\mathbf{s}(h,l)\Vert)|\Vert\mathbf{s}_h-\mathbf{s}_k\Vert\Vert\Delta\mathbf{s}(h,l)\Vert\nonumber\\
&+|\kappa(\Vert\Delta\mathbf{s}(k,l)\Vert)|\Vert\Delta\mathbf{s}(k,h)\Vert,
\end{align}
where $(a)$ follows from the triangle inequality, and $(b)$ holds because $\kappa(\Vert \Delta \mathbf{s}(h,l) \Vert)$ is a concave function of $\Vert \mathbf{s}(h,l) \Vert$. $(b)$ can also be understood from the physical meaning of $Z(k,l)$. When the SNR is low, or equivalently $\Vert \mathbf{s}(k,l) \Vert$ is small, $Z(k,l)$ is large and decreases as the SNR increases. Moreover, $Z(k,l)\ge 0$ makes it a monotonic function that decreases most rapidly at $\Vert \mathbf{s}(k,l) \Vert = 0$. For simplicity, define $V(h,l) = |\nabla_{\Vert \Delta \mathbf{s}(h,l) \Vert} \kappa(\Vert \Delta \mathbf{s}(h,l) \Vert)|\Vert \Delta \mathbf{s}(h,l) \Vert$. In this case, it follows that
\begin{align}
L(k,l)\geq&V(h,l)+|\kappa(\Vert\Delta\mathbf{s}(k,l)\Vert)|.
\end{align}

However, the first term in $L(k,l)$ is not determined since $\mathbf{s}_h$ represents an arbitrary feasible vector. Therefore, it is necessary to show that $V(h,l)$ is finite. For a fixed $\mathbf{s}_l$, the value of $\nabla_{\Vert \Delta \mathbf{s}(h,l) \Vert} \kappa(\Vert \Delta \mathbf{s}(h,l) \Vert)$ depends only on $\Vert \Delta \mathbf{s}(h,l) \Vert$. As previously described, $|\nabla_{\Vert \Delta \mathbf{s}(h,l) \Vert} \kappa(\Vert \Delta \mathbf{s}(h,l) \Vert)|$ attains its maximum at $\Vert \Delta \mathbf{s}(h,l) \Vert = 0$ and monotonically decreases to zero as $\Vert \Delta \mathbf{s}(h,l) \Vert \to +\infty$. Consequently, $V(h,l)$ is non-negative and vanishes as $\Vert \Delta \mathbf{s}(h,l) \Vert \to 0$ or $\Vert \Delta \mathbf{s}(h,l) \Vert \to +\infty$. Therefore, there exists a finite maximum at some finite $\Vert \Delta \mathbf{s}(h,l) \Vert$, and we have
\begin{align}
L(k,l)\geq|\kappa(\Vert\Delta\mathbf{s}(k,l)\Vert)|+\sup\limits_{\mathbf{s}_h\in\Omega_P(h)}V(h,l).
\end{align}

Note that the time index in $\Omega_P(h)$ can also be omitted. Clearly, $|\kappa(\Vert \Delta \mathbf{s}(k,l) \Vert)|$ is finite and thus, $L(k,l)$ is well-defined. Next, consider the case $\kappa(\Vert \Delta \mathbf{s}(h,l) \Vert) \le \kappa(\Vert \Delta \mathbf{s}(k,l) \Vert)$. Using similar manipulations, we can derive that
\begin{align}
&\Vert\kappa(\Vert\Delta\mathbf{s}(k,l)\Vert)\Delta\mathbf{s}(k,l)-\kappa(\Vert\Delta\mathbf{s}(h,l)\Vert)\Delta\mathbf{s}(h,l)\Vert\nonumber\\
\leq&(Z(\mathbf{s}_h)+V(k,l))\Vert\Delta\mathbf{s}(k,h)\Vert.
\end{align}

Combining the above analysis, the final global Lipschitz constant can be expressed as
\begin{align}
&L(k,l)\nonumber\\
=&\max\Big(\sup\limits_{\mathbf{s}_k\in\Omega_P(k),\mathbf{s}_l\in\Omega_P(l)}|\nabla_{\Vert\Delta\mathbf{s}(h,l)\Vert}\kappa(\Vert\Delta\mathbf{s}(h,l)\Vert)|,\nonumber\\
&\sup\limits_{\mathbf{s}_h\in\Omega_P(h)}Z(\mathbf{s}_h)\Big)+\sup\limits_{\mathbf{s}_k\in\Omega_P(k),\mathbf{s}_l\in\Omega_P(l)}V(h,l).
\end{align}

Finally, we can complete the proof based on the following inequalities for arbitrary $h$ and $l$,
\begin{align}
|\nabla_{\Vert\Delta\mathbf{s}(h,l)\Vert}\kappa(\Vert\Delta\mathbf{s}(h,l)\Vert)|\leq|\nabla\kappa(\mathbf{0})|,
\end{align}
\begin{align}
Z(\mathbf{s}_h)\leq Z(\mathbf{0}).
\end{align}
\end{appendices}

\footnotesize
\bibliographystyle{IEEEtran}
\bibliography{ref}

\end{document}